\documentclass[aps,pre,superscriptaddress,twocolumn,tightenlines,showpacs,floatfix,amsmath,amssymb,pdftex,usenames,dvipsnames]{revtex4-1}

\usepackage{amsmath}
\usepackage{amssymb}
\usepackage{bm}
\usepackage{bbm}
\usepackage{float}
\usepackage{tikz} 
\usepackage{graphicx}
\usepackage{subfigure}
\usepackage{multirow}
\usepackage{booktabs}
\usepackage[normalem]{ulem}
\usepackage{float} 
\def\id{\mathbbm{1}}

\newcommand{\corr}[1]{\langle #1\rangle}
\newcommand{\mb}[1]{\mathbf{#1}}
\newcommand{\mbb}[1]{\mathbb{#1}}
\newcommand{\mr}[1]{\mathrm{#1}}
\newcommand{\tmb}[1]{\widetilde{\mathbf{#1}}}

\begin{document}
\title{Generic first-order phase transitions between isotropic and orientational phases with polyhedral symmetries}
\date{\today}

\author{Ke Liu}
\affiliation{Arnold Sommerfeld Center for Theoretical Physics,
University of Munich, Theresienstrasse 37, 80333 Munich, Germany}

\author{Jonas Greitemann}
\affiliation{Arnold Sommerfeld Center for Theoretical Physics,
University of Munich, Theresienstrasse 37, 80333 Munich, Germany}

\author{Lode Pollet}
\affiliation{Arnold Sommerfeld Center for Theoretical Physics,
University of Munich, Theresienstrasse 37, 80333 Munich, Germany}

\begin{abstract}
Polyhedral nematics are examples of exotic orientational phases that possess a complex internal symmetry, representing highly non-trivial ways of rotational symmetry breaking, and are subject to current experimental pursuits in colloidal and molecular systems.
The classification of these phases has been known for a long time, however, their transitions to the disordered isotropic liquid phase remain largely unexplored, except for a few symmetries.
In this work, we utilize a recently introduced non-Abelian gauge theory to explore the nature of the underlying nematic-isotropic transition for all three-dimensional polyhedral nematics. 
The gauge theory can readily be applied to nematic phases with an arbitrary point-group symmetry, including those where traditional Landau methods and the associated lattice models may become too involved to implement owing to a prohibitive order-parameter tensor of high rank or (the absence of) mirror symmetries.
By means of exhaustive Monte Carlo simulations, we find that the nematic-isotropic transition is generically first-order for all polyhedral symmetries.
Moreover, we show that this universal result is fully consistent with our expectation from a renormalization group approach, as well as with other lattice models for symmetries already studied in the literature.
We argue that extreme fine tuning is required to promote those transitions to second order ones.
We also comment on the nature of phase transitions breaking the $O(3)$ symmetry in general cases.
\end{abstract}

\maketitle

\section{Introduction}
Nematic liquid crystal phases are states of matter that possess long-range orientational order but are translationally invariant \cite{DeGennesBook}. 
Historically, they were discovered in systems of rod-like molecules with a
$D_{\infty h}$ symmetry and had a revolutionary impact on the display industry.
However, it is generally accepted that the classification of nematic phases coincides
with three-dimensional ($3D$) point groups. Since the early $1970$s there
have been steady and tremendous efforts in the search for new nematic phases beyond uniaxial order.
Indeed, $D_{2h}$ biaxial nematics were proposed \cite{Freiser70} and their properties were discussed \cite{Alben73, Straley74} just shortly after theories of the uniaxial ones were established \cite{DeGennesBook}.
There is now strong evidence of their existence
\cite{Madsen04, Acharya04, Severing04} and they are believed to be
promising candidates for the next generation of liquid crystal displays (LCDs) \cite{BiaxialBook15}.
Another remarkable example of unconventional nematic phases whose existence has been established is the twist-bent liquid crystals formed from bent-core constituents with $C_{2v}$ symmetry \cite{Madsen04, Acharya04}.
They exhibit intriguing optical \cite{Merkel04, Neupane06, Gortz09, Chen14} and elastic \cite{Gleeson14, Kaur16} properties and rich transition sequences \cite{Lubensky02, Mettout05, Takezoe06, LongaPajakWydro09}, and are still subject to present studies.

Moreover, there is also great interest in more complex polyhedral
nematic phases classified by polyhedral groups. Attention has been focused on
the search for those phases experimentally in various chemical and colloidal systems \cite{Wiant08, Qazi10, Aoshima12, Lekkerkerker12} and numerically from packing shapes \cite{Blaak04, John04, John08, Duncan09, Duncan11, Marechal12, Wilson12, Glotzer12, Glotzer16a, Glotzer16b}, classifying the associated order parameter tensors \cite{Fel95, Michel01,HajiAkbariGlotzer15} and topological defects \cite{Mermin79, Michel80}, constructing theories supporting such orders and investigating their phase diagrams \cite{NelsonToner81, SteinhardtNelson1981, Lubensky02, Romano06, Romano08, TrojanowskiLonga12}, and examining the consequential macroscopic properties \cite{Stallinga94, Brand05}.

Nevertheless, in spite of considerable progress, we may have unveiled only a small corner of the rich landscape of the polyhedral phases. There are still many open questions on, e.g., their transition sequences and the nature of those phase transitions, the interactions of the associated topological defects and their influence on the thermodynamical, optical and mechanical properties of the system.
From a theoretical point of view, the difficulty is closely tied to the
complexity of those symmetries and their subgroup structure. These demand tensor
order parameters of high rank and lead to rich patterns of phase transitions where dynamics of topological defects also plays a crucial role.
Traditional Landau schemes and the associated lattice models are explicitly
based on order parameters, and, hence, may become extremely involved and difficult to handle when dealing with complicated symmetries. 
They are also not convenient in accessing topological defects.
Furthermore, the full classification of the explicit form of those order
parameter tensors has been attained only recently \cite{Liu16, Nissinen16}.

However, lattice gauge theory, adopted from high-energy physics \cite{Kogut79},
has opened up new avenues to address these issues.
The application of this method to nematic orders dates back to the seminal works
of Lammert, Rokshar, and Toner in the mid-$1990$s \cite{LammertRoksharToner93, LammertRoksharToner95, TonerLammertRokshar95b}. 
The authors utilized a $Z_2$ gauge theory to promote Heisenberg vectors to
directors, and formulated their model in terms of vectors and $Z_2$ gauge
fields, instead of $Q_{ab}$ tensors. They successfully capture the important
statistical physics of uniaxial nematics, especially the first-order
nematic-isotropic (NI) transition, 
and show the power of lattice gauge theory in controlling dynamics of topological defeccts.
Variants of the Lammert-Rokshar-Toner model have also been applied to strongly correlated electron systems, for instance, in studies of charge fractionalization of superconductors \cite{SenthilFisher00, SenthilFisher01, DemlerSenthil05} and spin nematics \cite{ZaanenNussinov02, PodolskyDemler05}.

The works mentioned above have focussed exclusively on $Z_2$
symmetries and uniaxial orders. Only recently has the gauge-theoretical description been extended to accommodate general point-group symmetries, in the studies of $2+1 d$ quantum melting \cite{Liu15, Beekman17} and $3D$ thermal nematics \cite{Liu16, Nissinen16, Liu17} by ourselves and collaborators.

Its advantages have been proven both mathematically and practically, especially when dealing with $3D$ point groups which are in general non-Abelian.
First of all, it has been shown in solid mathematical terms that the gauge model fits all nematic orders into a uniform and efficient framework, regardless of their symmetry \cite{Liu16}.
This is in stark contrast to traditional order-parameter methods which are typically specific only for a single symmetry and often suffer from the complexity arising from high-rank ordering tensors.
Moreover, the formulation of the gauge model requires no prior knowledge of the underlying order parameter which is an essential input for traditional methods. 
Instead, it acts as a machinery and generates a full classification of nematic ordering tensors which, to our knowledge, has never been done before \cite{Nissinen16}, though remarkable results of a more narrow scope have been obtained previously by other means \cite{Fel95, Michel01,HajiAkbariGlotzer15}.
Furthermore, in virtue of its generality, the gauge-theoretical method can also provide a global view over symmetries, which allows us to explore universal properties of different nematic orders \cite{Liu16}.
These include the insight of a relation between thermal fluctuations and symmetries, and the finding of a vestigial chiral phase that is reminiscent the chiral liquid reported in a recent experiment \cite{Dressel14}.
Last but not least, the gauge model is also naturally compatible with anisotropic interactions. By allowing anisotropy, it has mathematically predicted and numerically verified rich patterns of biaxial-uniaxial transitions and new types of biaxial-biaxial$^{\star}$ transitions  \cite{Liu17}, enriching our understanding of biaxial orders.

In earlier works, we have focused on building the connection between generalized $3D$ nematics and non-Abelian gauge theories, and on exploring the topology of their phase diagrams.
In the present paper we study the nature of the NI transition for polyhedral orders by means of Monte Carlo simulations and a renormalization group analysis. This is not only important for physical properties of the system near the phase transition, but also relates to a fundamental question in statistical physics. That is, whether breaking $O(3)$ in different manners can give rise to new universality classes.
Moreover, it is worth pointing out that the gauge-theory model allows us to
easily exclude irrelevant symmetries and focus on the most important degrees of
freedom. Meanwhile, the model remains flexible enough to incorporate competing orders and disorders.

The symmetries we are interested in are the $7$ polyhedral subgroups of $O(3)$, $\{T, T_d, T_h, O, O_h, I, I_h\}$, requiring orientational tensors of rank higher than $2$.
Nematic phases of these symmetries are sometimes dubbed as octupolar or tetrahedral ($T_d$), cubic ($O, O_h$) or icosahedral ($I, I_h$) phases in literature. However, for convenience we will refer to them as generalized nematic phases when discussing general symmetries and $G-$nematics when discussing a specific instance of the symmetry $G$.
This convention was by no means invented by us, but already used in the textbook Ref.~\onlinecite{DeGennesBook}.
These polyhedral nematic phases have not been clearly identified in experiments.
Nevertheless, this does not mean that they are only of academical interest.
Indeed, modern technologies in nano and colloid science are able to synthesize and
manipulate mesoscopic particles with the desired symmetry to a high degree
\cite{Glotzer07, Glotzer10, Kraft09, Kraft11, Mark13, Huang15}, hence providing essential building blocks for the realization of polyhedral phases.
Moreover, it is also worth noting that these phases may emerge from systems of lower-symmetry constituents with suitable interactions or geometrical constraints, such as the proposed tetrahedral $T_d$ phase from $C_{2v}$-shaped molecules \cite{Lubensky02, TrojanowskiLonga12} and the cubic $O_h$ phase from $D_{\infty h}$-components \cite{Blaak99, Marechal12, Romano16}.

This paper is organized as follows. In Section \ref{sec:model}, we define the necessary degrees of freedom, and review the realization of generalized nematics in the language of lattice gauge theories.
Section \ref{sec:results} is devoted to Monte Carlo simulations. We first discuss the results of the chiral tetrahedral $T$ nematics in detail, then present those for other polyhedral symmetries with a discussion on their general features. 
We compare our results with those from a renormalization scenario and other lattice models in Section \ref{sec:other_methods}.
Finally, we conclude and provide an outlook in Section \ref{sec:done}.

\section{Gauge-theory description of nematic phases} \label{sec:model} 

\subsection{Degrees of freedom} \label{subsec:dof}
Instead of directly using physical order parameters, the fundamental degrees
of freedom are nonphysical matter fields and gauge fields in the gauge theoretical description.
The matter fields are $O(3)$ rotors describing all possible rotations in $3D$ real space. They can be parameterized by local orthonormal triads as 
\begin{equation}
R = (\mb{l}, \mb{m}, \mb{n} )^{\mr{T}} \in O(3),
\end{equation}
where $\mb{n}^{\alpha} = \{\mb{l}, \mb{m}, \mb{n}\}$ with $\alpha = 1, 2, 3$ are the three axes of a local triad.
In concrete terms, they are defined by rotations that let $R$ coincide with the fixed ``laboratory'' axes $\mb{e}^{a} = \{\mb{e}_1, \mb{e}_2, \mb{e}_3 )$, and are parametrized by three Euler angles with respect to $\mb{e}^{a}$.

The three axes of $R$ satisfy the relation 
\begin{equation}
\sigma = \mr{det}(R) = \mb{l} \cdot (\mb{m} \times \mb{n}) = \pm 1. \label{eq:handedness}
\end{equation}
This also defines the chirality or handedness, denoted by a pseudo-scalar $\sigma$, of the triad.
For $\sigma = 1$, $R$ describes rotations in $SO(3)$ and is usually referred to
as a proper rotation.
For convenience later on, we also define triads formed by pseudo-vectors 
$\tmb{n}^{\alpha} = \{\tmb{l}, \tmb{m}, \tmb{n}\}$, 
\begin{equation}
\widetilde{R} = (\tmb{l}, \tmb{m}, \tmb{n} )^{\mr{T}} \in SO(3), \label{SO3_triad}
\end{equation}
with $\tmb{l} \cdot (\tmb{m} \times \tmb{n}) \equiv 1$, describing rotations of a rigid body.
Correspondingly, a rotation in $O(3)$ can be decomposed as
\begin{equation}
R = \sigma \widetilde{R}.
\end{equation}

The gauge fields are defined as a connection between two neighboring triads.
They are also rotations, but in contrast to the global $O(3)$ rotations, they describe local rotations with respect to the axes of a triad.
The introduction of gauge fields makes it is possible to compare two triads
locally at different locations. The symmetry of the gauge fields is a point
group by construction. 
In the simplest situation, when homogenous distributions of order parameters are preferred, it coincides with the
symmetry of the ``mesogens'' of a liquid crystal.
In terms of the terminology of Ref. \onlinecite{Mettout06}, this symmetry is
chosen to be the symmetry of the effective building blocks of the system, and
can in turn represent the symmetry of the state in the fully symmetry-broken
phase. In other words, the scheme works at a coarse-grained level by construction.

\subsection{The model} \label{subsec:model}

\begin{figure}
\centering
 \subfigure[]{
  \includegraphics[width=0.45\textwidth]{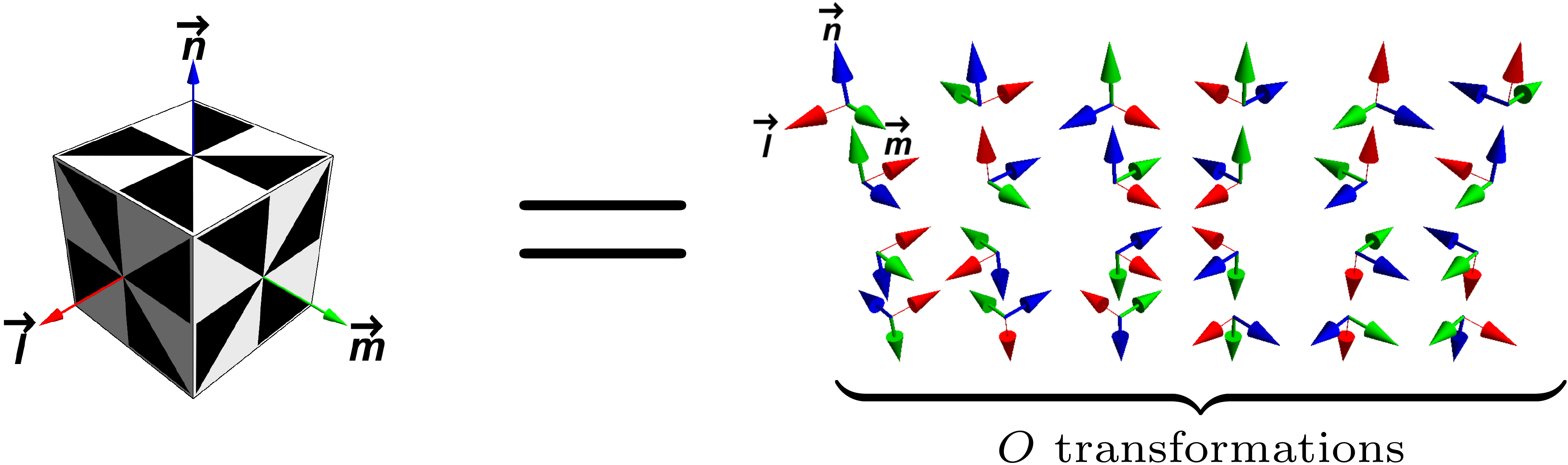}\label{fig:cube}
 }
 \vfill
 \subfigure[]{
  \includegraphics[width=0.45\textwidth]{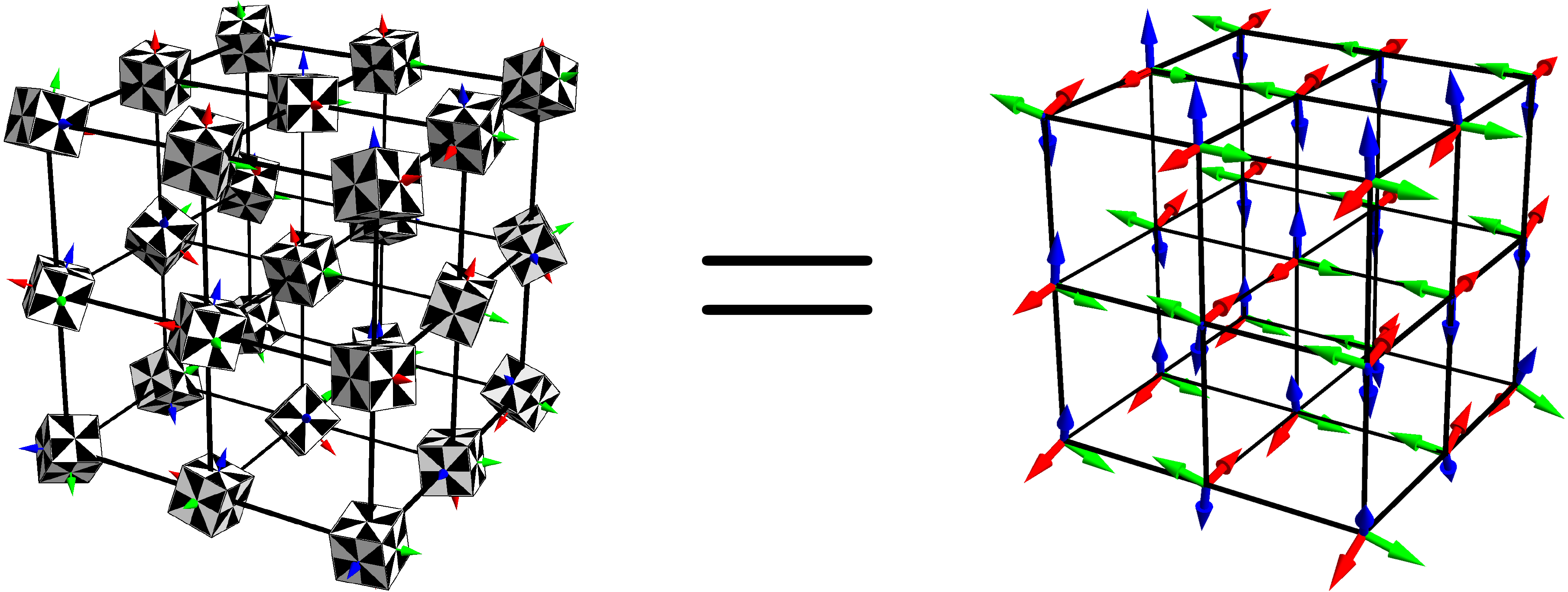}\label{fig:lattice}
 }
 
 \caption{ The correspondence between the shape of mesogens and the configurations of local triads. Here, a mesogen with the chiral cubic $O$ symmetry is considered as an example.
 Fig.~\ref{fig:cube}: The orientation of a cube, representing physical order parameter fields, is encoded into $24$
triad configurations, generated by a gauge symmetry $O$ acting on local axes of a triad.
Fig.~\ref{fig:lattice}: A lattice of cubes maps on to a lattice with triad matter
fields residing on the sites and point-group symmetric gauge fields defined on the links.
 }
\end{figure}

Having established the necessary information on the degrees of freedom, we now introduce the gauge-theoretic model. 
It is defined on an auxiliary cubic lattice, which is permitted by the continuous translational symmetry of nematic liquid crystals. It hence applies to both continuous and discrete point groups. 
The Hamiltonian takes the following form,
\begin{equation}
H = -\sum_{\corr{i,j}} \mr{Tr}(R_i^{\mr{T}} \mathbb{J} U_{ij} R_j). \label{eq:model}
\end{equation}
$R_i$ is the $O(3)$ triad at a lattice site $i$, $U_{ij}$ is the gauge field of a point-group symmetry $G$ mediating the interaction between two nearest-neighboring sites and lives on the link $\corr{i,j}$. 
$\mathbb{J}$ is a coupling matrix and can act as a tuning parameter. It is constrained by the symmetry of nematic mesogens, \textit{i.e.} the gauge symmetry $G$, in such a way that it has to be invariant under the transformation $\Lambda \mathbb{J} \Lambda^{\mr{T}} = \mathbb{J}$, $\forall \Lambda \in G$.
It follows that $\mathbb{J}$ is isotropic and takes the form $\mathbb{J} = J \id$ for all polyhedral groups, where $J$ is positive for ferromagnetic (alignment) coupling.
However, anisotropies of $\mbb{J}$ are possible for nematics with axial symmetries, and are responsible to the generalized biaxial-uniaxial and biaxial-biaxial$^{\star}$ transitions \cite{Liu17}.

The Hamiltonian Eq. \eqref{eq:model} is invariant under gauge transformations 
\begin{equation}
R_i \rightarrow \Lambda_i R_i,\quad U_{ij} \rightarrow \Lambda_i U_{ij} \Lambda_j^{\mr{T}},\quad \forall \Lambda_i \in G. \label{eq:gauge_trans}
\end{equation}
This invariance identifies the orientation of a triad defined by $R_i$ with that defined by $\Lambda_i R_{i}$, and thus encodes the symmetry of the mesogens under consideration.

To be more concrete, we define a local triad $n^{\prime \beta}_i = U^{\beta\gamma}_{ij} n^{\gamma}_j$ at a site $i$, and rewrite the Hamiltonian Eq. \eqref{eq:model} in the following form,
\begin{equation}
H = -\sum_{\corr{i,j}} J^{\alpha\beta} n^{\alpha}_i \cdot n^{\prime \beta}_j\label{eq:triad_form},
\end{equation}
where Greek letters in superscripts are associated with the axes of local triads, and $J^{\alpha\beta} = J \delta_{\alpha, \beta}$ for polyhedral symmetries.
By doing so, the triad $n^{\gamma}_j$ has been brought to the same local gauge, namely the same body-fixed coordinate, as the site $i$ by parallel transporting so that the orientation of the two triads can be compared.
Then, considering the gauge transformations in Eq. \eqref{eq:gauge_trans}
running over $G$ at a site $j$, but letting other sites unchanged, $\Lambda_{i
  \neq j} \equiv \id$, this generates a set of $n^{\prime \beta}_j$ which
consists of all the equivalent definitions of the orientation of the underlying mesogen of the symmetry $G$ at the site $j$.
Let us take a chiral cube with the symmetry $O$ as an example. The orientation of the cube maps to $24$ configurations of a local triad, corresponding to the $24$ transformations of the group $O$.
When all these configurations are considered and identified, we are effectively describing the orientation of a cube via that of a set of local triads. The symmetry of the underlying mesogens is thus realized by the gauge symmetry. Note that the choice of $\Lambda_{i \neq j} \equiv \id$ in the above example is purely for simplifying the example. 
Gauge transformations Eq. \eqref{eq:gauge_trans} can be performed independently to all the sites.
Consequently, in the low-energy limit, the orientational interaction of physical mesogens (order parameter fields) is hence effectively encoded in the gauge model Eq. \eqref{eq:model} of nonphysical degrees of freedom, as depicted by Fig. \ref{fig:lattice}.
This procedure can also be shown in explicit mathematical terms by integrating out the gauge fields Eq. \eqref{eq:model}, and we refer to our earlier publications Refs. \onlinecite{Liu16} and \onlinecite{Nissinen16} for detailed proofs.
However, though maybe less intuitive, it is advantageous to work with gauge degrees of freedom.
As the symmetry of the order parameter fields is directly implemented by the gauge symmetry, the gauge model applies to all point group symmetries by simply choosing a desired $G$.

\subsection{Discussion on the phases}
It is well known that gauge symmetries cannot break spontaneously \cite{Elitzur75}. As a consequence, the fully symmetry-broken phase of the gauge model, Eq. \eqref{eq:model}, features a ground state manifold $O(3)/G$ which is just the order-parameter manifold of a $G-$nematic phase.
This phase is usually referred to as the Higgs phase in the language of gauge theories, and corresponds to the aligned state of mesogens in the current context, i.e., a nematic phase of the symmetry $G$.
On the other hand, the disordered $O(3)$ liquid phase is realized by the confinement phase of the gauge model Eq. \eqref{eq:model}.

There are three comments to be made on the above statements to avoid confusion.
First, the Higgs phase just mentioned corresponds to a situation where $O(3)$
has been completely broken to the local symmetry $G$. However, aside from this,
there can also be an intermediate Higgs phase that breaks $O(3)$ to a larger point group $G^{\prime}$ satisfying $G \subset G^{\prime} \subset O(3)$, featuring vestigial order.
This could happen when fluctuations in some sectors of the degrees of freedom
are more pronounced than in others. 
For instance, in case $G$ is a finite axial group, the fully ordered Higgs phase is a biaxial nematic phase of the symmetry $G$. 
When fluctuations in the plane perpendicular to the so-called primary axis are sufficiently strong or weak, upon changing temperature, the system may experience an intermediate uniaxial and/or biaxial$^{\star}$ phase, respectively, before entering the disordered isotropic liquid phase, as discussed in detail in Ref. \onlinecite{Liu17}.
As another example, an intermediate Higgs phase may also appear as a chiral
liquid phase. Possible realizations of this phase are systems formed from
mesogens of a chiral polyhedral symmetry, $G \in \{T, O, I\}$. For these
symmetries, fluctuations in orientations are much more pronounced than those in the chirality. 
Thus a phase that breaks real-space inversion and mirror symmetry but is invariant under $SO(3)$ rotations can emerge between the nematic phase and the $O(3)$ liquid phase. We will encounter this situation again in the next section and a systematic and detailed discussion can be found in Ref. \onlinecite{Liu16}.

Second, the distinction between the Higgs phase and the confinement phase is only a property when $G$ is a nontrivial subgroup of $O(3)$. In the limit $G = O(3)$, these two phases are continuously connected and indistinguishable \cite{FradkinShenker79}, consistent with the fact that there is no symmetry breaking for $O(3)/O(3)$.

Last but no least, as mentioned earlier, we will focus on homogeneous distributions of order parameter fields, as is realized by the gauge model Eq. \eqref{eq:model}.
However, inhomogeneous distributions may also lead to interesting phenomena. One example is the chiral-$T$ phase with a helical structure of $T_d$ molecules, owing to an explicit chiral elastic term in Landau free energy, as discussed in Ref.~\onlinecite{Fel95}. We can also introduce a gauge invariant chiral term to Eq.~\eqref{eq:model} to incorporate such helical structure for general symmetries, but will leave it for future study.
What is relevant to the current paper is that, as such a chiral term is independent to the additional quartic terms of high rank tensors, it is unlikely that they can change the nature of fluctuation-induced first-order phase transitions for the $T$ and $T_d$ symmetry which will be discussed in Sec. \ref{subsec:MF_FG}.

 \subsection{Topological defects}
Before closing the section, let us briefly comment on the dynamics of gauge fields in the model Eq. \eqref{eq:model} and its relation to topological defects of liquid crystals.

From the point of view of gauge theories, the model Eq. \eqref{eq:model} consists of a single Higgs term, in which form the dynamics of gauge fields $U_{ij}$ arises purely from the interaction with the matter fields $R_i$.
In general, however, the gauge fields can have their own dynamics, which in the simplest case is described by a plaquette term in the following form,
\begin{equation}
H_{\mr{YM}} = - \sum_{\square} K_{C_{\mu}} (U_{\square}) \mr{Tr}[U_{\square}]. \label{eq:K_term}
\end{equation}
This generalizes the defect suppression term in  Refs.~\onlinecite{LammertRoksharToner93, LammertRoksharToner95}, and essentially is
 an analog of the Yang-Mills theory as the $U_{ij}$'s are in general non-Abelian \cite{Kogut79}. However, comparing to usual lattice Yang-Mills theories in high energy physics, here we are interested in discrete symmetries.
$U_{\square} \in G$ denotes the orientated product of the gauge fields around a minimal plaquette of the lattice, $U_{\square} = \prod_{\corr{ij} \in \square} U_{ij}$. A plaquette of $U_{\square} \neq \id$ is frustrated and represents a gauge flux. 
The coupling strength $K_{C_{\mu}}$ depends on the trace of $U_{\square}$, so it is a function of the conjugation class $C_{\mu}$ of $G$, which means that gauge fluxes in the same conjugation class are physically equivalent.

A gauge flux has the effect that after a triad travels around it in a closed circuit, the triad is rotated by $U_{\square}$, just like circling a disclination. Furthermore, the classification of gauge fluxes coincides with the Volterra classification of disclinations \cite{Kleman08}. Even though this classification, as well as the Volterra classification, is in general not identical to the homotopy classification of topologically stable defects \cite{Mermin79}, it includes all the elementary topologically stable defects and can be used to construct the full homotopy classification.
As we can easily tune $K_{C_{\mu}}$ to suppress or prompt certain types of defects, the interaction in $H_{\mr{YM}}$ provides a possible route to study the influence of topological defects on thermodynamical properties of nematic liquid crystals.
As is known from the study of lattice gauge theory \cite{FradkinShenker79}, this may qualitatively change both the topology of the phase diagrams and the nature of underlying phase transitions. 
	Refs.~\onlinecite{LammertRoksharToner93, LammertRoksharToner95, TonerLammertRokshar95b} also showed remarkable examples in this context, where the first-order uniaxial-isotropic transition is split into two continuous ones when the defect suppression is sufficiently large.
 For general symmetries, we expect rich physics to explore when treating the elastic term ~\eqref{eq:model} and non-Abelian defect term ~\eqref{eq:K_term} at equal footing.
Nevertheless, this is beyond the scope of the current paper, and deserves a separate systematic study. Thus, for simplicity, we will set $K_{C_{\mu}} \equiv 0$ and focus on Eq. \eqref{eq:model} in the following. Physically, this means none of the topological defects are assigned a particular core energy.

\setlength{\tabcolsep}{2pt}
\renewcommand{\arraystretch}{1.8}
\begin{table}
\centering
\caption{{\bf Generators of $3D$ polyhedral point groups.} 
The first column specifies the symmetries. The second column shows a set of generators which produce the entire group elements of the underlying symmetry. The third column gives the order of the symmetries. Note that, there are multiple ways to choose the generator set, but they are all equivalent (for more information see Refs. \onlinecite{BookButler12} and \onlinecite{BookBishop93}). A representation of the generators listed below is catalogued in Table \ref{table:generators}.
} 
\label{table:groups}
\begin{tabular}{ | c | c | c | }
    \hline
    \hline
   \parbox{2.1 cm}{\centering{\bf Symmetry}} 
	& \parbox{3.5 cm}{\centering{\bf Generators}}  
	& \parbox{1.3 cm}{\centering{\bf Order}}  
   \\ \hline

    {\centering{$T$}}
    & $c_2(\mathbf{n})$, $c_3(\mathbf{l}+\mathbf{m}+\mathbf{n})$	
	& 12	
   \\ \hline

    {\centering{$T_d$}}
    & $c_2(\mathbf{n})$, $c_3(\mathbf{l}+\mathbf{m}+\mathbf{n})$	, $\sigma_d$
	& 24
	\\ \hline   
	
	{\centering{$T_h$}}
    & $c_2(\mathbf{n})$, $c_3(\mathbf{l}+\mathbf{m}+\mathbf{n})$	, $\sigma_h$
	& 24
	\\ \hline
	
	{\centering{$O$}}
    & $c_2(\mathbf{m} + \mathbf{n})$, 
    	$c_3(\mathbf{l}+\mathbf{m}+\mathbf{n})$, $c_4({\mathbf{n}})$ 
	& 24	
   \\ \hline
   
   	{\centering{$O_h$}}
    & \parbox{3.5 cm}{\centering $c_2(\mathbf{m} + \mathbf{n})$, 
    	$c_3(\mathbf{l}+\mathbf{m}+\mathbf{n})$, $c_4({\mathbf{n}})$, $\sigma_h$}
	& 48	
   \\ \hline
   
   	{\centering{$I$}}
    & $c_2(\mathbf{n})$, $c_3(\mathbf{l}+\mathbf{m}+\mathbf{n})$, $c_5(\mathbf{l} + \tau\mathbf{n})$ 
	& 60	
   \\ \hline
   
     {\centering{$I_h$}}
    & \parbox{3.5 cm} {\centering $c_2(\mathbf{n})$, $c_3(\mathbf{l}+\mathbf{m}+\mathbf{n})$, $c_5(\mathbf{l} + \tau\mathbf{n})$, $\sigma_h$}
    & 120	
   \\ \hline
  
   \hline 
\end{tabular}

\end{table}

\setlength{\tabcolsep}{2pt}
\renewcommand{\arraystretch}{1.8}
\begin{table}
\centering
\caption{{\bf Definitions of the generators.}
Here we specify the representation of the generators for the $3D$ polyhedral point groups used in our simulations. $c_N(\mathbf{p})$ denotes a $N$-fold rotation about a vector $\mathbf{p}$ defined by the local axes $\{\mathbf{l}, \mathbf{m}, \mathbf{n}\}$. $\tau = (\sqrt{5}+1)/2$ is the golden ratio, which is involved in case of icosahedral groups $I$ and $I_h$. $\sigma_h$ defines a reflection about the $(\mathbf{l}, \mathbf{m})$ plane, while $\sigma_d$ indicates a reflection about the plane $(\mathbf{l} + \mathbf{m}, \mathbf{n})$.
}
\label{table:generators}
\begin{tabular}{ | c | c | }
    \hline
    \hline
   \parbox{2.9 cm}{\centering{\bf Generator}} 
	& \parbox{5 cm}{\centering{\bf Representation}}  
   \\ \hline

	{\centering $c_{2}(\mathbf{n})$}   
	& $\begin{pmatrix}
 	-1  & 0 & 0 \\
 	0 & -1 & 0 \\
 	0 & 0 & 1
	\end{pmatrix}$ 
  \\ \hline
  
   {\centering $c_{2}(\mathbf{m}+\mathbf{n})$}   
	& $\begin{pmatrix}
	 -1 & 0 & 0 \\
	 0 & 0 & 1 \\
	 0 & 1 & 0
	\end{pmatrix}$ 
  \\ \hline
  
  	{\centering $c_{3}(\mathbf{l}+\mathbf{m}+\mathbf{n})$}   
	& $\begin{pmatrix}
	 0 & 1 & 0 \\
	 0 & 0 & 1 \\
	 1 & 0 & 0
	\end{pmatrix}$ 
  \\ \hline
  
 	{\centering $c_{4}(\mathbf{n})$}   
	& $\begin{pmatrix}
 	0  & -1 & 0 \\
 	1 & 0 & 0 \\
 	0 & 0 & 1
	\end{pmatrix}$ 
  \\ \hline 
  
  	{\centering $c_{5}(\mathbf{l} + \tau \mathbf{n})$}   
	& $\begin{pmatrix}
 	1/2 & -\tau/2 & 1/(2\tau) \\
 	\tau/2 & 1/(2\tau) & -1/2 \\
 	1/(2\tau) & 1/2 & \tau/2
	\end{pmatrix}$ 	
  \\ \hline
  
     {\centering $\sigma_h$}   
	& $\begin{pmatrix}
	 1 & 0 & 0 \\
	 0 & 1 & 0 \\
	 0 & 0 & -1
	\end{pmatrix}$ 
  \\ \hline
  
      {\centering $\sigma_d$}   
	& $\begin{pmatrix}
	 0 & -1 & 0 \\
	 -1 & 0 & 0 \\
	 0 & 0 & 1
	\end{pmatrix}$ 
  \\ \hline

   \hline 
\end{tabular}

\end{table}

\section{Numerical results} \label{sec:results}

As discussed in the last section, the gauge model Eq. \eqref{eq:model} can act
as an efficient and flexible framework for studying nematic order with arbitrary point-group symmetries. Moreover, it is readily accessible by Monte Carlo simulations.
In light of this, we examined the nature of the nematic-isotropic transition for all $3D$ polyhedral groups, \textit{i.e.}, $\{T, T_d, T_h, O, O_h, I, I_h\}$. 
For the convenience of the reader, generators of these symmetries are provided in Tables \ref{table:groups} and \ref{table:generators}, while more information can be found in textbooks, e.g., Refs. \onlinecite{BookButler12} and \onlinecite{BookBishop93}. Sch\"{o}nflies notation is used thorough out the manuscript.
We use the Metropolis algorithm and run simulations on cubic lattices with volume $V = 8^3, 16^3, 24^3$.

The simulations include three steps. In the first step, the transition
temperatures are located as precise as possible by examining the peak position of the heat capacity and the nematic susceptibility (see below). 
Procedures of cooling random initial states and heating uniform states are compared.
In the next step, we perform extensive simulations near the transition, histogramming the distribution of the observables of interest. The typical amount of \textit{independent} samples used are of the order of $10^5$ to $10^7$.
In the last step, the histograms are further improved by Ferrenberg-Swendsen reweighting \cite{Ferrenberg88, BookNewman99}, and the transition temperatures are estimated from the shape of the histograms \cite{Lee90, BookLandau13}.  
 
In the rest of the section, we first present the results for tetrahedral $T$ nematics and discuss the general features of this phase transition in detail. Then we discuss other symmetries.

\subsection{$SO(3)/T$ transition} \label{subsec:SO3_T} 

\begin{figure}\centering
  \includegraphics[width=0.45\textwidth]{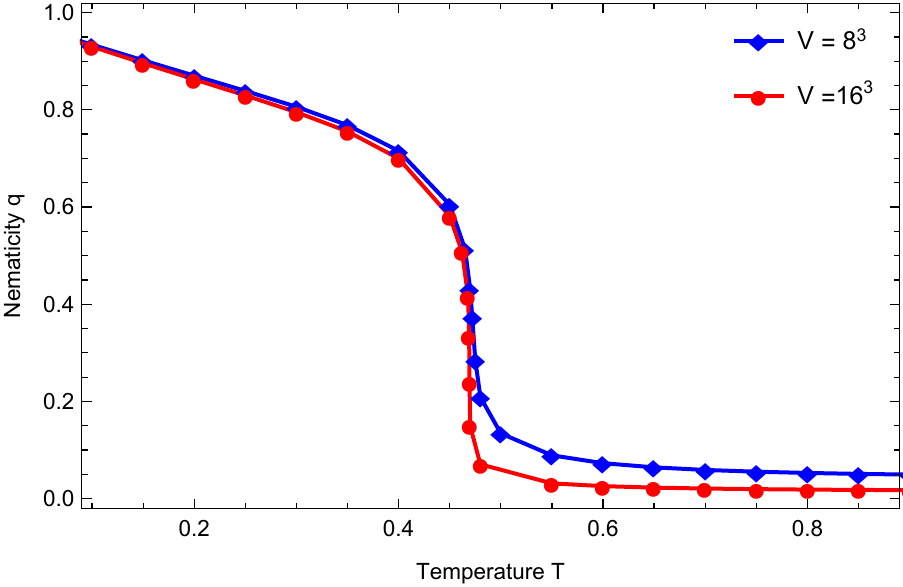} 
  \caption{ Nematicity of the tetrahedral $T$ symmetry for a broad range of temperatures in units of the Higgs coupling $J$, computed by Eq. \eqref{eq:SO3}. Error bars are smaller than the data points. These, together with the susceptibility and heat capacity (not shown), give us information of the nontrivial region where massive simulations are performed to locate and determine the order of the transition.} \label{fig:Nematicity}
\end{figure}

The tetrahedral group $T$  consists of $12$ proper rotations leaving a tetrahedron invariant. Therefore, aside from the orientational order, a $T$-symmetric nematic phase also takes an intrinsic chiral order, breaks inversion and any kinds of mirror symmetries of real space.
Note that the $T$-nematic phase we discuss here is different from the $T$-phase discussed by Fel in Ref. \onlinecite{Fel95}.
In the later case the $T$ symmetry arises from the helical structure of $T_d$-symmetric mesogens and is associated with a different order parameter (see below and also Sec. \ref{subsec:MF_FG} for further discussion).

It turns out that for nematics formed from constituents with a $T$-symmetry and
a flexible handedness, as well as from those with an $O$- or $I$-symmetry,
fluctuations in the orientation sector are in general much more pronounced than those in the chirality sector \cite{Liu16}. 
Consequently, the system develops orientational order and chiral order sequentially.
Furthermore, by comparing numerical results with a mean-field analysis, it is shown that the two phase transitions are well separated.
This implies that the relevant degrees of freedom associated with the NI transition lie in the $SO(3)$ sector in Eq. \eqref{eq:model}.

In mathematical terms, we can rewrite the gauge model Eq. \eqref{eq:model} as
\begin{equation}
H = - J \sum_{\corr{i,j}} \sigma_i \sigma_j \mr{Tr}(\widetilde{R}_i^{\mr{T}} U_{ij} \widetilde{R}_j), 
\end{equation}
by taking out the $Z_2$ center of $O(3)$, where $\sigma_i$ denotes the handedness fields and $\widetilde{R}_i$ are $SO(3)$ triads, defined in Eq. \eqref{eq:handedness} and Eq. \eqref{SO3_triad}.
The handedness fields are frozen during the NI transition, thus simplifying the problem to a phase transition breaking $SO(3)$, governed by a Hamiltonian 
\begin{equation}
H^{\prime} = - J \sum_{\corr{i,j}} \mr{Tr}(\widetilde{R}_i^{\mr{T}} U_{ij} \widetilde{R}_j). \label{eq:SO3}
\end{equation}

\begin{figure}
\centering
 \subfigure[]{
  \includegraphics[width=0.45\textwidth]{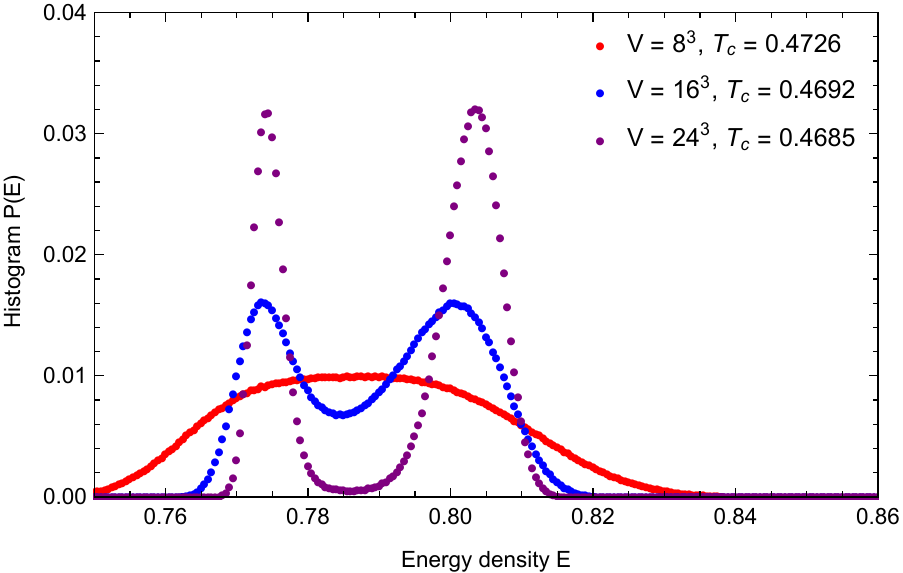}\label{fig:Histo_E_T}
 }
 \vfill
 \subfigure[]{
  \includegraphics[width=0.45\textwidth]{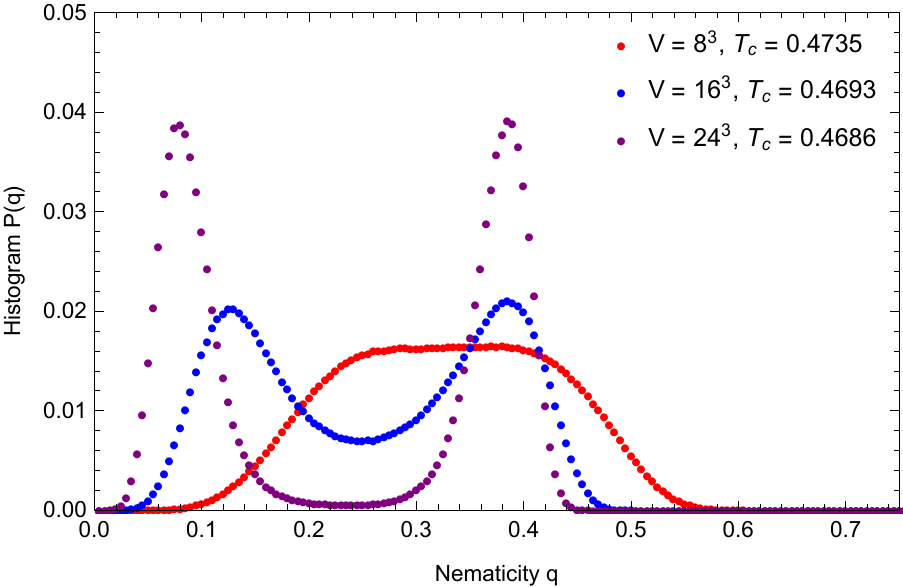}\label{fig:Histo_Nem_T}
 }
 
 \caption{ Histograms of the energy density, Fig. \ref{fig:Histo_E_T}, and of the nematicity, Fig, \ref{fig:Histo_Nem_T}, for tetrahedral $T$ nematics at the phase transition at different lattice sizes. Transition temperatures, $T_c$, are estimated by finding where the histogram peaks are of equal height. The data shown have been improved by Ferrenberg-Swendsen reweighting in the vicinity of the simulated temperatures, which are at the precision of four decimal digits.}\label{fig:Histo_T}
\end{figure} 

The orientational order parameter of $T$-nematic phases is a rank-$3$ tensor taking the following form (note the difference to the $T_d$ order parameter, see Sec. \ref{subsec:MF_FG}),
\begin{align} \label{eq:op_T}
\overline{\mbb{O}}^{T} &= \corr{\mbb{O}^{T}_i}_V \nonumber \\
&= \sum_{\mr{cyc}}\corr{\tmb{l} \otimes \tmb{m} \otimes \tmb{n} - \frac{1}{6}\varepsilon_{abc} \tmb{e}_a \otimes \tmb{e}_b \otimes \tmb{e}_c }_V,
\end{align}
where $\mbb{O}^{T}_i$ denotes the local ordering tensor at a coarse-grained lattice site $i$, $\corr{...}_V$ denotes the average over the volume, and $\sum_{\mr{cyc}}$ is the sum running over cyclic permutations of local axes $\{\tmb{l}, \tmb{m}, \tmb{n}\}$ \cite{Nissinen16}.
The Levi-Civita tensor is introduced to make the ordering tensor traceless, so $\overline{\mbb{O}}^{T}$ becomes a zero tensor in the liquid phase. This term is only needed when working with $SO(3)$ triads where the handedness of each local triad is fixed. If the handedness is allowed to fluctuate (\textit{i.e}, the case of an $O(3)$ triad) summing over the two kinds of chirality cancels this term.

In case of homogenous distributions, instead of using the tensor form Eq. \eqref{eq:op_T}, we can characterize a nematic order of symmetry $G$ by its magnitude defined as
\begin{equation}
q = \sqrt{\mr{Tr}(\mbb{O}^G \cdot \mbb{O}^G)}.
\end{equation}
This quantity is called the  \textit{nematicity} and generalizes the concept of magnetization \cite{Nissinen16}.
Consequently, we can further define the susceptibility of $q$ in  the standard way,
\begin{equation}
\chi_{q} = \beta V \left( \corr{q^2} - \corr{q}^2 \right),
\end{equation}
and detect the NI transition by the peak of $\chi_{q}$, where $\beta$ is the inverse temperature.
As confirmed in our simulations, the peak of $\chi_{q}$ coincides with that of the heat capacity defined in the standard way, indicating that the nematicity $q$ is indeed a valid scalar order parameter.

We have measured the $SO(3)/T$ NI transition with Eq. \eqref{eq:SO3} by monitoring several quantities, including the energy, the nematicity, histograms of the two, the heat capacity and the susceptibility. As many of them reveal the same information, we present only those which are necessary for the following discussions. 
In Fig. \ref{fig:Nematicity}, we show the behavior of the nematicity for a broad range of temperatures, and Figs. \ref{fig:Histo_E_T} and \ref{fig:Histo_Nem_T} show the histograms of the energy density and the  nematicity at phase transition, $P(E)$ and $P(q)$, respectively.

As shown in the energy histogram Fig. \ref{fig:Histo_E_T}, a double-peak
behavior emerges at sufficiently large lattice sizes, and appears more pronounced when the lattice size increases, indicating the occurrence two stable, co-existing phases, which is a hallmark of a first-order phase transition. The distance between the valley and peak of a histogram, measuring the difficulty for the system to tunnel between two phases, also increases with the lattice size, as expected.
The physical meaning of the two peaks is revealed by the nematicity histogram
from the same simulations, shown in Fig. \ref{fig:Histo_Nem_T}. With increasing
lattice size, aside from the behavior that the peaks become more pronounced, the
first peak notably moves to the left. We expect it eventually goes to zero in the thermodynamical limit, indicating a disordered liquid phase.  The other peak corresponds to the nematic phase.


\subsection{Other symmetries} \label{subsec:others}

\begin{figure}
\centering
  \includegraphics[width=0.45\textwidth]{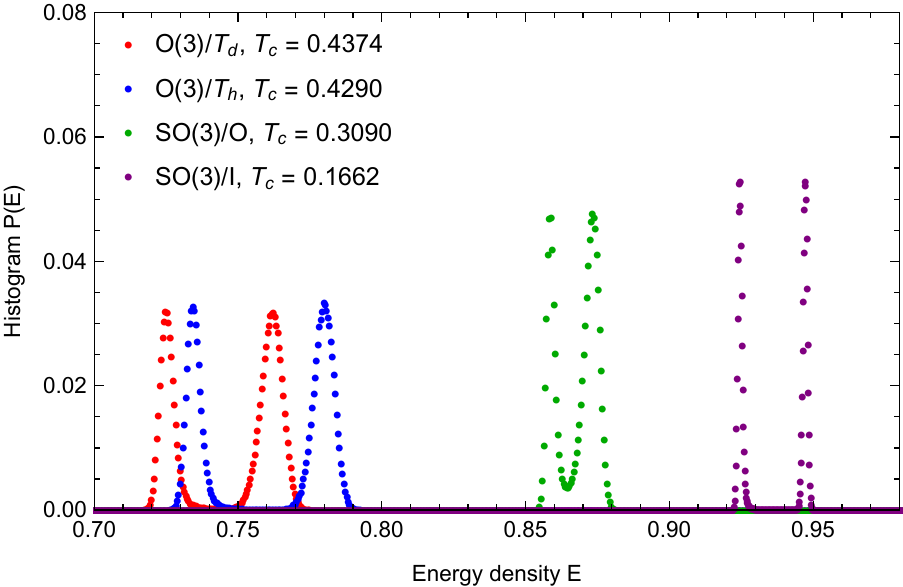} 
  \caption{ Histograms of the energy density for symmetries $\{T_d, T_h, O, I\}$ at the NI transition, for lattice size $V = 24^3$. $T_c$'s are estimated in the same way as for the $SO(3)/T$ case. 
Depending on symmetries, the simulations are performed with the original gauge theory Eq. \eqref{eq:model} or its $SO(3)$ sector Eq. \eqref{eq:SO3}, indicated in the figure. The data of the $SO(3)/O$ and $SO(3)/I$ transition also represent those of the $O(3)/O_h$ and $O(3)/I_h$ transition, respectively. 
The two-peaks behavior reveals the first-order nature for all these transitions. The histograms of the nematicity (not shown) exhibit similar features.
The same binning size is used for the first three symmetries, while a smaller
binning size is used for the $SO(3)/I$ transition. However, the heights of these
histograms are not comparable, since even though these symmetries are studied
via a common framework here, they correspond to different physical models. 
   } \label{fig:Histo_others}
\end{figure} 

We performed the same procedure as in the case of the $SO(3)/T$ NI transition for all other polyhedral symmetries, \textit{i.e.}, symmetries of $\{T_d, T_h, O, O_h, I, I_h\}$. Similar to the $T$-nematics, the breaking of chiral and of rotational symmetry in case of the cubic $O$ and the icosahedral $I$ symmetry are separated, owing to huge fluctuations in orientation \cite{Liu16}.
Thus, they are also simulated in terms of Eq. \eqref{eq:SO3}, resulting in a $SO(3)/O$ and a $SO(3)/I$ transition, respectively.
On the other hand, the nonchiral symmetries $\{T_d, T_h, O_h, I_h\}$ are studied via the original model Eq. \eqref{eq:model}, corresponding to a direct breaking of $O(3)$.

We find first-order behavior for all these transitions. Fig. \ref{fig:Histo_others} shows the energy histogram of symmetries $\{T_h, T_d, O, I\}$.
The results of the $SO(3)/O$ and $SO(3)/I$ transition also represent those of the $O(3)/O_h$ and $O(3)/I_h$ transition, which have very similar behavior as the former two, with slightly higher $T_c$'s.
This may be understood by the fact that the $Z_2$ center in the latter two cases can be factorized as a trivial $Z_2/Z_2$ theory, leading to the same order parameter manifold as the $SO(3)/O$ and $SO(3)/I$ cases, respectively. 
Indeed, $O$-nematics and $O_h$ nematics, as well as $I$- and $I_h$-nematics,
share the same orientational order parameter, only distinguished by a pseudo-scalar chiral order parameter \cite{Nissinen16}.  

We also studied the behavior of the nematicity and its histogram for these symmetries. Although the curves corresponding to different symmetries are well separated (since the phase transitions occur at different temperature scales) in Fig. \ref{fig:Histo_others},
the histograms of the nematicities  overlap closely for these symmetries,
especially in the disorder region where all the disordering peaks are located at some small nematicity value close to $0$. 
Moreover, they show very similar features as seen in Figs. \ref{fig:Nematicity} and \ref{fig:Histo_T} for the $SO(3)/T$ transition case.
They are therefore not presented. 

One notable feature of Fig. \ref{fig:Histo_others} is that the peaks of the
histogram shift to lower energy scales as symmetries increase (the energy density is normalized via $E_G = - 9V$ which is the energy when all mesogens uniformly align up),
indicating a decrease in the corresponding transition temperatures. This can be understood as
a consequence of more pronounced orientational fluctuations for high symmetries, which in turn results in an increasing difficulty to stabilize the order.
Note that this feature is manifest when using a common metric, the Higgs coupling strength $J$, for all the symmetries. This metric is not a direct physical measure in the sense that it describes the interaction strength between the auxiliary gauge fields and matter fields, rather than that of physical order parameter fields. 
Although the latter one depends in principle on the Higgs coupling, the derivation of the relation is in general nontrivial. Nevertheless, this does not prevent us to obtain general insights on the nature of the phase transitions.
Clearly, Fig. \ref{fig:Histo_others} reveals the generic first-order nature of
the NI transition for all these symmetries. It is not clear yet how the strength
of these first-order transitions depends on their respective symmetries. However, this depends on microscopic details of a system and a universal conclusion might not exist.

\section{Relations and comparisons to other methods} \label{sec:other_methods}

We have numerically reached the conclusion of a generic first order NI transition from a particular framework. Moreover, this is consistent with existing results from other methods, including a general perspective from mean field theories, renormalization group (RG) analyses \cite{Radzihovsky01} and other lattice models \cite{Romano06, Romano08}, as we elaborate below.

\subsection{Mean-field theories and RG} \label{subsec:MF_FG}

A significant difference between nematic order and spin or vector order is that in general the former requires a  tensor order parameter due to nontrivial internal symmetries. In case of the $D_{\infty h}$ uniaxial nematics, the order parameter is a rank-$2$ tensor, $Q_{ab}$, which gives rise to a third order term, $Q_{ab} Q_{bc} Q_{ca}$, in the Landau-de Gennes theory and makes the NI transition discontinuous.

For polyhedral symmetries $\{T_h, O, O_h, I, I_h\}$ the nematic order parameters are also even rank tensors \cite{Nissinen16}, which take the following form for nonchiral groups $\{T_h, O_h, I_h\}$,
\begin{align} 
\mathbb{O}^{T_h} 
&=  \mathbf{l}^{\otimes 2} \otimes \mathbf{m}^{\otimes 2} +  \mathbf{m}^{\otimes 2} \otimes \mathbf{n}^{\otimes 2} +  \mathbf{n}^{\otimes 2} \otimes \mathbf{l}^{\otimes 2}
\nonumber \\
 & \quad -\frac{2}{5} \delta_{ab} \delta_{cd}\bigotimes_{\substack{ \mu =  a,b,c, d}} \mathbf{e}_{\mu}
+ \frac{1}{10} \bigg( \delta_{ac} \delta_{bd} \bigotimes_{\substack{ \mu =  a,c,b, d}} \mathbf{e}_{\mu}
\nonumber \\   
& \quad    + \delta_{ad} \delta_{bc}\bigotimes_{\substack{ \mu =  a,d,b, c}} \mathbf{e}_{\mu} \bigg), \label{eq:op_Th}\\
\mathbb{O}^{O_h} &=  \mathbf{l}^{\otimes 4}  +  \mathbf{m}^{\otimes 4} + \mathbf{n}^{\otimes 4} 
-\frac{1}{5}\sum_{\text{pairs}} \delta_{ab} \delta_{cd} \bigotimes_{\substack{ \mu =  a,b,c, d}} \mathbf{e}_{\mu}, \label{eq:op_Oh} \\
\mathbb{O}^{I_h} &= \sum_{\rm{cyc}} \bigg[ \mathbf{l}^{\otimes 6} +\sum_{ \{+,- \}} 
			\Big(\frac{1}{2} \mathbf{l} \pm \frac{\tau}{2} \mathbf{m} \pm \frac{1}{2\tau} \mathbf{n} \Big)^{\otimes 6} \bigg]
\nonumber \\		
  		& \quad -\frac{1}{7} \sum_{\text{pairs}} \delta_{ab} \delta_{cd} \delta_{ef} \bigotimes_{\substack{ \mu =  a,b,c, \\  d,e,f}} \mathbf{e}_{\mu}, \label{eq:op_Ih}
\end{align}
and $\mathbb{O}^{O} = \{\mathbb{O}^{O_h}, \sigma\}$ and $\mathbb{O}^{I} = \{\mathbb{O}^{I_h}, \sigma\}$ for chiral groups $\{O, I\}$, where $\sum_{\text{cyc}}$ runs over the cyclic permutations of $\{\mb{l}, \mb{m}, \mb{n}\}$, $\sum_{\text{pairs}}$ sums over all nonequivalent pairings of the indices of the Kronecker delta functions, and $\otimes$ denotes the tensor product.
Similar to the $D_{\infty h}$ uniaxial case, even in naive mean-field theories for these symmetries, there are third order terms of the form 
$\mbb{O}_{abcd} \mbb{O}_{cdef} \mbb{O}_{efab} $
or $\mbb{O}_{abcdef} \mbb{O}_{defghk} \mbb{O}_{ghkabc}$,
giving rise to a first order phase transition.

On the other hand, the tetrahedral $T_d$ order requires a rank-$3$ order parameter tensor, $\mbb{O}^{T_d}$, where
\begin{align} \label{eq:op_Td}
\mathbb{O}^{T_d} =  \sum_{\mathrm{cyc}}  (\mathbf{l} \otimes \mathbf{m} +  \mathbf{m} \otimes \mathbf{l})\otimes \mathbf{n}.
\end{align}
This forbids the appearance of the third order term in a naive mean-field theory which has the following free energy density,
\begin{align} \label{eq:MF_Td}
f_{T_d} = \frac{1}{2} (\partial_i \mbb{O}^{T_d}_{abc} \partial_i \mbb{O}^{T_d}_{abc}) + \frac{r}{2} \mbb{O}^{T_d}_{abc} \mbb{O}^{T_d}_{abc} + u (\mbb{O}^{T_d}_{abc} \mbb{O}^{T_d}_{abc})^2,
\end{align}
and predicts a continuous NI transition, where $r$ and $u$ are phenomenological coefficients.

A further RG study by Radzihovsky and Lubensky in Ref. \onlinecite{Radzihovsky01} shows that Eq. \eqref{eq:MF_Td} has a second order phase transitions, falling into the $O(7)$ universality class, as the symmetric traceless tensor $\mbb{O}^{T_d}$ can be mapped to a $7$ dimensional vector.
However, this transition is unstable against fluctuations. Following the general symmetry principles there exists another fourth order term, 
$\mbb{O}^{T_d}_{abc} \mbb{O}^{T_d}_{ade} \mbb{O}^{T_d}_{bdf} \mbb{O}^{T_d}_{cef}$, 
representing fluctuations, which qualitatively modifies the nature of the NI transition and makes it first order.

The tetrahedral $T$ order faces a similar situation as the $T_d$ one. Its order parameter $\mbb{O}^T$ defined in Eq. \eqref{eq:op_T} is also an odd rank and symmetric traceless tensor (The kernel of $\mbb{O}^T$ is not symmetric, but becomes so after carrying out the trace). Therefore, the paradigm of Ref. \onlinecite{Radzihovsky01} for the $T_d$ case equally applies to the $SO(3)/T$ transition, with a different tensor representation for the $O(7)$ vector. Consequently, the second order NI transition predicted by the MF theory is converted to a first order one, agreeing with our results from the gauge model.

With this new perspective, let us look back to the symmetries $\{T_h, O, O_h, I, I_h\}$. Although the first-order nature of their NI transitions can already be concluded from a naive mean-field treatment, we should not omit other quartic couplings for completeness. 
$\mbb{O}^{O_h}$ and $\mbb{O}^{I_h}$ are also symmetric and traceless tensors, and lead to $5$ and $7$ quartic terms 
(the number of independent terms may be reduced by one), respectively.
It is more tricky in the $T_h$ case. $\mbb{O}^{T_h}$ is only partially symmetric (it is only invariant under switching the first or the last two indices), hence even the second and the third order coupling are not unique.
Of course not all of those couplings are necessary to appear in a particular
system.  However, this implies that it requires very precise fine tuning to
promote the NI transition for polyhedral symmetries to second order.

\subsection{Other lattice models} \label{subsec:other_lattice}
Next we compare our results with those from other lattice models. Remarkable examples are Refs. \onlinecite{Romano06} and \onlinecite{Romano08} for the $T_d$ and $O_h$ order. (Other examples have been mainly focused on uniaxial and biaxial orders \cite{BiaxialBook15}.)
They are typically constructed by an interaction potential between two rigid molecules or mesogens of a certain symmetry.
The orientation of a mesogen is described by $M$ (nonorthogonal) unit vectors, spanned from a common local origin and organized in a way that explicitly has the desired symmetry.

The potential, $V_{ij}$, is then defined in terms of Legendre polynomials of the inner product of those vectors at the lowest nontrivial order,
\begin{align} \label{eq:pair_potential}
	V_{ij} &= V(\mb{v}^{(m)}_i, \mb{v}^{(m^{\prime})}_j) \nonumber \\
	& \sim -c_l \sum^M_{m,m^{\prime} = 1} P_l(\mb{v}^{(m)}_i \cdot \mb{v}^{(m^{\prime})}_j),
\end{align} 
where the coefficient $c_l$ is positive for ferromagnetic coupling, $P_l(...)$ denotes the Legendre polynomial at the order $l$, and $\mb{v}^{(m)}_i$ are local unit vectors at the lattice site $i$.

For the cubic $O_h$ order, the $\mb{v}^{(m)}_i$ coincide with our local triads
$\mb{n}^{\alpha}_i$, and the Legendre polynomials are trivial for $l < 4$
\cite{Romano06}. On the other hand, in case of the tetrahedral $T_d$ order,
the $\mb{v}^{(m)}_i$ are the $4$ three-fold axes of a regular tetrahedron and are
nonorthogonal \cite{Fel95}, whereas the Legendre polynomials become nontrivial only
from $P_3$ onwards \cite{Romano08}.
Following the same principle and given the expression Eq. \eqref{eq:op_Ih}, one expects that this requires $15$ vectors, which are the $2$-fold axes of a regular icosahedron \cite{Fel95}, and sixth order Legendre polynomials.

It is not hard to show that the interaction potential Eq. \eqref{eq:pair_potential} can be understood as a counterpart of the Lebwohl-Lasher model \cite{Lebwohl72} for general symmetries, and is equivalent to the inner product between two order parameter tensors, 
\begin{equation} \label{eq:effective_potential}
	V_{ij} \sim \mr{Tr}(\mbb{O}^G_i \cdot \mbb{O}^G_j).
\end{equation}
As shown in Refs. \onlinecite{Liu16, Nissinen16}, this is exactly the leading order of the effective Hamiltonian of the gauge model Eq. \eqref{eq:model} after tracing out the gauge fields.
Therefore, it is not a surprise that our results of the $T_d$ and $O_h$ NI transition agree with those in Refs. \onlinecite{Romano06, Romano08}, and the agreements to other polyhedral symmetries, $\{T, T_h, O, I, I_h\}$, can also be expected.
However, lattice models for these symmetries in the type of the potential Eq.
\eqref{eq:pair_potential} are yet to be developed to the best of our knowledge.
Unlike the gauge model Eq. \eqref{eq:model}, which can readily be applied for all point-group symmetries, the potential Eq. \eqref{eq:pair_potential} is symmetry-dependent and involves large amounts of vectors and high-order Legendre polynomials, whose high complexity in actual use can be anticipated.
For instance, in case of the $I_h$ order, it involves $225$ Legendre polynomials of order $6$.

However, the potential Eq. \eqref{eq:pair_potential} has advantages in a more straightforward connection with microscopic interactions of liquid crystal mesogens, as it is built directly on physical order parameter fields.
Moreover, it is interesting to see how this method applies to the $T$ and $T_h$
symmetries, where the role of mirrors may be manifest, as well as the relation and difference of the resultant lattice models to those of the $T_d$ and $O_h$ case, which are the symmetry they halve, respectively.

\section{Summary} \label{sec:done}

Rotational symmetry breaking is ubiquitous and plays an important role in
condensed matter physics and statistical physics. One of its intriguing features
is that there is a multitude of ways to break a symmetry into its subgroups, leading to a large array of exotic phases.
In this work, we have examined the nature of phase transitions breaking the rotational group $O(3)$ to polyhedral point groups.
Such phases  are  prime candidates  in the search for unconventional nematic liquid crystals, in particular in the field of nano and colloidal science.

We found that the transitions from the nematic phase to the isotropic liquid phase are generically first order for all polyhedral symmetries.
Furthermore, the polyhedral NI transitions are robust in the sense that they require fine tuning of a high precision in order to achieve a second order phase transition. This feature is inherited from the complexity of the group structure of polyhedral symmetries.

Moreover, along the lines of the discussion in Sec. \ref{subsec:MF_FG}, we
anticipate the NI transition of generalized uni- and bi-axial nematics, which
breaks $O(3)$ to axial point groups $\{C_n, C_{nv}, C_{nh}, S_{2n}, D_{n},
D_{nh}, D_{nd} \}$, to be generically of first order as well.
As discussed in detail in Ref. \cite{Nissinen16}, the order parameter of
axial symmetries in general has the structure $\mbb{O}^G = \{\mbb{A}^G,
\mbb{B}^G, \sigma\}$, where $\mbb{A}^G = \mbb{A}^G[\mb{n}]$ defines the order of
the primary axis chosen to be $\mb{n}$, $\mbb{B}^G = \mbb{B}^G[\mb{l}, \mb{m}]$
or $\mbb{B}^G[\mb{l}, \mb{m}, \mb{n}]$ defines the order in the perpendicular
plane and is required for finite axial symmetries, and $\sigma$ defines the
chiral order as seen in Sec. \ref{subsec:SO3_T} and is only relevant for the proper axial groups $\{C_n, D_n\}$.
For symmetries $\{ C_{nh}, S_{2n}, D_{n}, D_{nh}, D_{nd} \}$, $\mbb{A}^G$ is a
rank-two tensor and coincides with the $Q_{ab}$ director. Hence, following the
Landau-de Gennes theory, it is immediately clear that regardless of the in-plane structure, the NI transition for these symmetries will be generically first order.
For symmetries $C_n$ and $C_{nv}$, the primary order parameter $\mbb{A}^G$ is a vector, and continuous phase transitions seem to be preferred. However, when $n > 1$ but finite, the direct NI transition will be also first order, owing to the existence of an even and/or high rank $\mbb{B}^G$ tensor, as in the cases of polyhedral symmetries. 
Even at $n = 1$, where both $\mbb{A}$ and $\mbb{B}$ are vectorial, the order of the phase transition will depend on their coupling.
Therefore, even though there are diverse patterns to break the $O(3)$ symmetry, second-order transitions and corresponding universality classes may be quite rare. 
The familiar Heisenberg universality class related to the breaking of $O(3)$ to $O(2) \cong C_{\infty v}$ is a special case.
Our results add new insights to the physics of exotic orientational phases
and hopefully facilitate the understanding of future experiments.


Finally, we would like to note that in the present work only a single symmetry is considered in the realization of each polyhedral nematic.
Nevertheless, as has been discussed by many authors for $T_d$ and $O_h$ symmetries \cite{Lubensky02, TrojanowskiLonga12, Blaak99, Marechal12, Romano16}, polyhedral phases may emerge from systems formed from less-symmetric constituents.
Although it is hard to imagine a second-order NI transition from this, it would be interesting to explore the general pattern of symmetry emergence in liquid crystal systems.
Given the compatibility with competing orders and the potential power on
controlling topological defects, we expect that the gauge-theory scenario will be suitable to achieve this aim without losing simplicity.

\textbf{Acknowledgments} 

We would like to thank Leo Radzihovsky and Henk Bl\"{o}te for stimulating discussions, and Jaakko Nissinen for useful discussions and related collaborations.
This work is supported by FP7/ERC starting grant No. 306897. Our simulations made use of the ALPSCore library \cite{Gaenko17}.

\bibliographystyle{apsrev4-1}
\bibliography{polyhedra}

\begin{thebibliography}{80}%
\makeatletter
\providecommand \@ifxundefined [1]{%
 \@ifx{#1\undefined}
}%
\providecommand \@ifnum [1]{%
 \ifnum #1\expandafter \@firstoftwo
 \else \expandafter \@secondoftwo
 \fi
}%
\providecommand \@ifx [1]{%
 \ifx #1\expandafter \@firstoftwo
 \else \expandafter \@secondoftwo
 \fi
}%
\providecommand \natexlab [1]{#1}%
\providecommand \enquote  [1]{``#1''}%
\providecommand \bibnamefont  [1]{#1}%
\providecommand \bibfnamefont [1]{#1}%
\providecommand \citenamefont [1]{#1}%
\providecommand \href@noop [0]{\@secondoftwo}%
\providecommand \href [0]{\begingroup \@sanitize@url \@href}%
\providecommand \@href[1]{\@@startlink{#1}\@@href}%
\providecommand \@@href[1]{\endgroup#1\@@endlink}%
\providecommand \@sanitize@url [0]{\catcode `\\12\catcode `\$12\catcode
  `\&12\catcode `\#12\catcode `\^12\catcode `\_12\catcode `\%12\relax}%
\providecommand \@@startlink[1]{}%
\providecommand \@@endlink[0]{}%
\providecommand \url  [0]{\begingroup\@sanitize@url \@url }%
\providecommand \@url [1]{\endgroup\@href {#1}{\urlprefix }}%
\providecommand \urlprefix  [0]{URL }%
\providecommand \Eprint [0]{\href }%
\providecommand \doibase [0]{http://dx.doi.org/}%
\providecommand \selectlanguage [0]{\@gobble}%
\providecommand \bibinfo  [0]{\@secondoftwo}%
\providecommand \bibfield  [0]{\@secondoftwo}%
\providecommand \translation [1]{[#1]}%
\providecommand \BibitemOpen [0]{}%
\providecommand \bibitemStop [0]{}%
\providecommand \bibitemNoStop [0]{.\EOS\space}%
\providecommand \EOS [0]{\spacefactor3000\relax}%
\providecommand \BibitemShut  [1]{\csname bibitem#1\endcsname}%
\let\auto@bib@innerbib\@empty
\bibitem [{\citenamefont {de~Gennes}\ and\ \citenamefont
  {Prost}(1995)}]{DeGennesBook}%
  \BibitemOpen
  \bibfield  {author} {\bibinfo {author} {\bibfnamefont {P.}~\bibnamefont
  {de~Gennes}}\ and\ \bibinfo {author} {\bibfnamefont {J.}~\bibnamefont
  {Prost}},\ }\href@noop {} {\emph {\bibinfo {title} {The Physics of Liquid
  Crystals}}},\ International Series of Monographs on Physics\ (\bibinfo
  {publisher} {Clarendon Press},\ \bibinfo {year} {1995})\BibitemShut {NoStop}%
\bibitem [{\citenamefont {Freiser}(1970)}]{Freiser70}%
  \BibitemOpen
  \bibfield  {author} {\bibinfo {author} {\bibfnamefont {M.~J.}\ \bibnamefont
  {Freiser}},\ }\href {\doibase 10.1103/PhysRevLett.24.1041} {\bibfield
  {journal} {\bibinfo  {journal} {Phys. Rev. Lett.}\ }\textbf {\bibinfo
  {volume} {24}},\ \bibinfo {pages} {1041} (\bibinfo {year}
  {1970})}\BibitemShut {NoStop}%
\bibitem [{\citenamefont {Alben}(1973)}]{Alben73}%
  \BibitemOpen
  \bibfield  {author} {\bibinfo {author} {\bibfnamefont {R.}~\bibnamefont
  {Alben}},\ }\href@noop {} {\bibfield  {journal} {\bibinfo  {journal}
  {Physical Review Letters}\ }\textbf {\bibinfo {volume} {30}},\ \bibinfo
  {pages} {778} (\bibinfo {year} {1973})}\BibitemShut {NoStop}%
\bibitem [{\citenamefont {Straley}(1974)}]{Straley74}%
  \BibitemOpen
  \bibfield  {author} {\bibinfo {author} {\bibfnamefont {J.~P.}\ \bibnamefont
  {Straley}},\ }\href@noop {} {\bibfield  {journal} {\bibinfo  {journal}
  {Physical Review A}\ }\textbf {\bibinfo {volume} {10}},\ \bibinfo {pages}
  {1881} (\bibinfo {year} {1974})}\BibitemShut {NoStop}%
\bibitem [{\citenamefont {Madsen}\ \emph {et~al.}(2004)\citenamefont {Madsen},
  \citenamefont {Dingemans}, \citenamefont {Nakata},\ and\ \citenamefont
  {Samulski}}]{Madsen04}%
  \BibitemOpen
  \bibfield  {author} {\bibinfo {author} {\bibfnamefont {L.~A.}\ \bibnamefont
  {Madsen}}, \bibinfo {author} {\bibfnamefont {T.~J.}\ \bibnamefont
  {Dingemans}}, \bibinfo {author} {\bibfnamefont {M.}~\bibnamefont {Nakata}}, \
  and\ \bibinfo {author} {\bibfnamefont {E.~T.}\ \bibnamefont {Samulski}},\
  }\href {\doibase 10.1103/PhysRevLett.92.145505} {\bibfield  {journal}
  {\bibinfo  {journal} {Phys. Rev. Lett.}\ }\textbf {\bibinfo {volume} {92}},\
  \bibinfo {pages} {145505} (\bibinfo {year} {2004})}\BibitemShut {NoStop}%
\bibitem [{\citenamefont {Acharya}\ \emph {et~al.}(2004)\citenamefont
  {Acharya}, \citenamefont {Primak},\ and\ \citenamefont {Kumar}}]{Acharya04}%
  \BibitemOpen
  \bibfield  {author} {\bibinfo {author} {\bibfnamefont {B.~R.}\ \bibnamefont
  {Acharya}}, \bibinfo {author} {\bibfnamefont {A.}~\bibnamefont {Primak}}, \
  and\ \bibinfo {author} {\bibfnamefont {S.}~\bibnamefont {Kumar}},\ }\href
  {\doibase 10.1103/PhysRevLett.92.145506} {\bibfield  {journal} {\bibinfo
  {journal} {Phys. Rev. Lett.}\ }\textbf {\bibinfo {volume} {92}},\ \bibinfo
  {pages} {145506} (\bibinfo {year} {2004})}\BibitemShut {NoStop}%
\bibitem [{\citenamefont {Severing}\ and\ \citenamefont
  {Saalw{\"a}chter}(2004)}]{Severing04}%
  \BibitemOpen
  \bibfield  {author} {\bibinfo {author} {\bibfnamefont {K.}~\bibnamefont
  {Severing}}\ and\ \bibinfo {author} {\bibfnamefont {K.}~\bibnamefont
  {Saalw{\"a}chter}},\ }\href@noop {} {\bibfield  {journal} {\bibinfo
  {journal} {Physical review letters}\ }\textbf {\bibinfo {volume} {92}},\
  \bibinfo {pages} {125501} (\bibinfo {year} {2004})}\BibitemShut {NoStop}%
\bibitem [{\citenamefont {Luckhurst}\ and\ \citenamefont
  {Sluckin}(2015)}]{BiaxialBook15}%
  \BibitemOpen
  \bibinfo {editor} {\bibfnamefont {G.~R.}\ \bibnamefont {Luckhurst}}\ and\
  \bibinfo {editor} {\bibfnamefont {T.~J.}\ \bibnamefont {Sluckin}},\ eds.,\
  \href@noop {} {\emph {\bibinfo {title} {Biaxial Nematic Liquid Crystals:
  Theory, Simulation and Experiment}}}\ (\bibinfo  {publisher} {John Wiley \&
  Sons},\ \bibinfo {year} {2015})\BibitemShut {NoStop}%
\bibitem [{\citenamefont {Merkel}\ \emph {et~al.}(2004)\citenamefont {Merkel},
  \citenamefont {Kocot}, \citenamefont {Vij}, \citenamefont {Korlacki},
  \citenamefont {Mehl},\ and\ \citenamefont {Meyer}}]{Merkel04}%
  \BibitemOpen
  \bibfield  {author} {\bibinfo {author} {\bibfnamefont {K.}~\bibnamefont
  {Merkel}}, \bibinfo {author} {\bibfnamefont {A.}~\bibnamefont {Kocot}},
  \bibinfo {author} {\bibfnamefont {J.~K.}\ \bibnamefont {Vij}}, \bibinfo
  {author} {\bibfnamefont {R.}~\bibnamefont {Korlacki}}, \bibinfo {author}
  {\bibfnamefont {G.~H.}\ \bibnamefont {Mehl}}, \ and\ \bibinfo {author}
  {\bibfnamefont {T.}~\bibnamefont {Meyer}},\ }\href {\doibase
  10.1103/PhysRevLett.93.237801} {\bibfield  {journal} {\bibinfo  {journal}
  {Phys. Rev. Lett.}\ }\textbf {\bibinfo {volume} {93}},\ \bibinfo {pages}
  {237801} (\bibinfo {year} {2004})}\BibitemShut {NoStop}%
\bibitem [{\citenamefont {Neupane}\ \emph {et~al.}(2006)\citenamefont
  {Neupane}, \citenamefont {Kang}, \citenamefont {Sharma}, \citenamefont
  {Carney}, \citenamefont {Meyer}, \citenamefont {Mehl}, \citenamefont
  {Allender}, \citenamefont {Kumar},\ and\ \citenamefont {Sprunt}}]{Neupane06}%
  \BibitemOpen
  \bibfield  {author} {\bibinfo {author} {\bibfnamefont {K.}~\bibnamefont
  {Neupane}}, \bibinfo {author} {\bibfnamefont {S.~W.}\ \bibnamefont {Kang}},
  \bibinfo {author} {\bibfnamefont {S.}~\bibnamefont {Sharma}}, \bibinfo
  {author} {\bibfnamefont {D.}~\bibnamefont {Carney}}, \bibinfo {author}
  {\bibfnamefont {T.}~\bibnamefont {Meyer}}, \bibinfo {author} {\bibfnamefont
  {G.~H.}\ \bibnamefont {Mehl}}, \bibinfo {author} {\bibfnamefont {D.~W.}\
  \bibnamefont {Allender}}, \bibinfo {author} {\bibfnamefont {S.}~\bibnamefont
  {Kumar}}, \ and\ \bibinfo {author} {\bibfnamefont {S.}~\bibnamefont
  {Sprunt}},\ }\href {\doibase 10.1103/PhysRevLett.97.207802} {\bibfield
  {journal} {\bibinfo  {journal} {Phys. Rev. Lett.}\ }\textbf {\bibinfo
  {volume} {97}},\ \bibinfo {pages} {207802} (\bibinfo {year}
  {2006})}\BibitemShut {NoStop}%
\bibitem [{\citenamefont {Gortz}\ \emph {et~al.}(2009)\citenamefont {Gortz},
  \citenamefont {Southern}, \citenamefont {Roberts}, \citenamefont {Gleeson},\
  and\ \citenamefont {Goodby}}]{Gortz09}%
  \BibitemOpen
  \bibfield  {author} {\bibinfo {author} {\bibfnamefont {V.}~\bibnamefont
  {Gortz}}, \bibinfo {author} {\bibfnamefont {C.}~\bibnamefont {Southern}},
  \bibinfo {author} {\bibfnamefont {N.~W.}\ \bibnamefont {Roberts}}, \bibinfo
  {author} {\bibfnamefont {H.~F.}\ \bibnamefont {Gleeson}}, \ and\ \bibinfo
  {author} {\bibfnamefont {J.~W.}\ \bibnamefont {Goodby}},\ }\href {\doibase
  10.1039/B808283A} {\bibfield  {journal} {\bibinfo  {journal} {Soft Matter}\
  }\textbf {\bibinfo {volume} {5}},\ \bibinfo {pages} {463} (\bibinfo {year}
  {2009})}\BibitemShut {NoStop}%
\bibitem [{\citenamefont {Chen}\ \emph {et~al.}(2014)\citenamefont {Chen},
  \citenamefont {Nakata}, \citenamefont {Shao}, \citenamefont {Tuchband},
  \citenamefont {Shuai}, \citenamefont {Baumeister}, \citenamefont {Weissflog},
  \citenamefont {Walba}, \citenamefont {Glaser}, \citenamefont {Maclennan},\
  and\ \citenamefont {Clark}}]{Chen14}%
  \BibitemOpen
  \bibfield  {author} {\bibinfo {author} {\bibfnamefont {D.}~\bibnamefont
  {Chen}}, \bibinfo {author} {\bibfnamefont {M.}~\bibnamefont {Nakata}},
  \bibinfo {author} {\bibfnamefont {R.}~\bibnamefont {Shao}}, \bibinfo {author}
  {\bibfnamefont {M.~R.}\ \bibnamefont {Tuchband}}, \bibinfo {author}
  {\bibfnamefont {M.}~\bibnamefont {Shuai}}, \bibinfo {author} {\bibfnamefont
  {U.}~\bibnamefont {Baumeister}}, \bibinfo {author} {\bibfnamefont
  {W.}~\bibnamefont {Weissflog}}, \bibinfo {author} {\bibfnamefont {D.~M.}\
  \bibnamefont {Walba}}, \bibinfo {author} {\bibfnamefont {M.~A.}\ \bibnamefont
  {Glaser}}, \bibinfo {author} {\bibfnamefont {J.~E.}\ \bibnamefont
  {Maclennan}}, \ and\ \bibinfo {author} {\bibfnamefont {N.~A.}\ \bibnamefont
  {Clark}},\ }\href {\doibase 10.1103/PhysRevE.89.022506} {\bibfield  {journal}
  {\bibinfo  {journal} {Phys. Rev. E}\ }\textbf {\bibinfo {volume} {89}},\
  \bibinfo {pages} {022506} (\bibinfo {year} {2014})}\BibitemShut {NoStop}%
\bibitem [{\citenamefont {Gleeson}\ \emph {et~al.}(2014)\citenamefont
  {Gleeson}, \citenamefont {Kaur}, \citenamefont {G{\"o}rtz}, \citenamefont
  {Belaissaoui}, \citenamefont {Cowling},\ and\ \citenamefont
  {Goodby}}]{Gleeson14}%
  \BibitemOpen
  \bibfield  {author} {\bibinfo {author} {\bibfnamefont {H.~F.}\ \bibnamefont
  {Gleeson}}, \bibinfo {author} {\bibfnamefont {S.}~\bibnamefont {Kaur}},
  \bibinfo {author} {\bibfnamefont {V.}~\bibnamefont {G{\"o}rtz}}, \bibinfo
  {author} {\bibfnamefont {A.}~\bibnamefont {Belaissaoui}}, \bibinfo {author}
  {\bibfnamefont {S.}~\bibnamefont {Cowling}}, \ and\ \bibinfo {author}
  {\bibfnamefont {J.~W.}\ \bibnamefont {Goodby}},\ }\href@noop {} {\bibfield
  {journal} {\bibinfo  {journal} {ChemPhysChem}\ }\textbf {\bibinfo {volume}
  {15}},\ \bibinfo {pages} {1251} (\bibinfo {year} {2014})}\BibitemShut
  {NoStop}%
\bibitem [{\citenamefont {Kaur}(2016)}]{Kaur16}%
  \BibitemOpen
  \bibfield  {author} {\bibinfo {author} {\bibfnamefont {S.}~\bibnamefont
  {Kaur}},\ }\href {\doibase 10.1080/02678292.2016.1232442} {\bibfield
  {journal} {\bibinfo  {journal} {Liquid Crystals}\ }\textbf {\bibinfo {volume}
  {43}},\ \bibinfo {pages} {2277} (\bibinfo {year} {2016})},\ \Eprint
  {http://arxiv.org/abs/http://dx.doi.org/10.1080/02678292.2016.1232442}
  {http://dx.doi.org/10.1080/02678292.2016.1232442} \BibitemShut {NoStop}%
\bibitem [{\citenamefont {Lubensky}\ and\ \citenamefont
  {Radzihovsky}(2002)}]{Lubensky02}%
  \BibitemOpen
  \bibfield  {author} {\bibinfo {author} {\bibfnamefont {T.~C.}\ \bibnamefont
  {Lubensky}}\ and\ \bibinfo {author} {\bibfnamefont {L.}~\bibnamefont
  {Radzihovsky}},\ }\href {\doibase 10.1103/PhysRevE.66.031704} {\bibfield
  {journal} {\bibinfo  {journal} {Phys. Rev. E}\ }\textbf {\bibinfo {volume}
  {66}},\ \bibinfo {pages} {031704} (\bibinfo {year} {2002})}\BibitemShut
  {NoStop}%
\bibitem [{\citenamefont {Mettout}(2005)}]{Mettout05}%
  \BibitemOpen
  \bibfield  {author} {\bibinfo {author} {\bibfnamefont {B.}~\bibnamefont
  {Mettout}},\ }\href {\doibase 10.1103/PhysRevE.72.031706} {\bibfield
  {journal} {\bibinfo  {journal} {Phys. Rev. E}\ }\textbf {\bibinfo {volume}
  {72}},\ \bibinfo {pages} {031706} (\bibinfo {year} {2005})}\BibitemShut
  {NoStop}%
\bibitem [{\citenamefont {Takezoe}\ and\ \citenamefont
  {Takanishi}(2006)}]{Takezoe06}%
  \BibitemOpen
  \bibfield  {author} {\bibinfo {author} {\bibfnamefont {H.}~\bibnamefont
  {Takezoe}}\ and\ \bibinfo {author} {\bibfnamefont {Y.}~\bibnamefont
  {Takanishi}},\ }\href {http://stacks.iop.org/1347-4065/45/i=2R/a=597}
  {\bibfield  {journal} {\bibinfo  {journal} {Japanese Journal of Applied
  Physics}\ }\textbf {\bibinfo {volume} {45}},\ \bibinfo {pages} {597}
  (\bibinfo {year} {2006})}\BibitemShut {NoStop}%
\bibitem [{\citenamefont {Longa}\ \emph {et~al.}(2009)\citenamefont {Longa},
  \citenamefont {Pajak},\ and\ \citenamefont {Wydro}}]{LongaPajakWydro09}%
  \BibitemOpen
  \bibfield  {author} {\bibinfo {author} {\bibfnamefont {L.}~\bibnamefont
  {Longa}}, \bibinfo {author} {\bibfnamefont {G.}~\bibnamefont {Pajak}}, \ and\
  \bibinfo {author} {\bibfnamefont {T.}~\bibnamefont {Wydro}},\ }\href@noop {}
  {\bibfield  {journal} {\bibinfo  {journal} {Physical Review E}\ }\textbf
  {\bibinfo {volume} {79}},\ \bibinfo {pages} {040701} (\bibinfo {year}
  {2009})}\BibitemShut {NoStop}%
\bibitem [{\citenamefont {Wiant}\ \emph {et~al.}(2008)\citenamefont {Wiant},
  \citenamefont {Neupane}, \citenamefont {Sharma}, \citenamefont {Gleeson},
  \citenamefont {Sprunt}, \citenamefont {J\'akli}, \citenamefont {Pradhan},\
  and\ \citenamefont {Iannacchione}}]{Wiant08}%
  \BibitemOpen
  \bibfield  {author} {\bibinfo {author} {\bibfnamefont {D.}~\bibnamefont
  {Wiant}}, \bibinfo {author} {\bibfnamefont {K.}~\bibnamefont {Neupane}},
  \bibinfo {author} {\bibfnamefont {S.}~\bibnamefont {Sharma}}, \bibinfo
  {author} {\bibfnamefont {J.~T.}\ \bibnamefont {Gleeson}}, \bibinfo {author}
  {\bibfnamefont {S.}~\bibnamefont {Sprunt}}, \bibinfo {author} {\bibfnamefont
  {A.}~\bibnamefont {J\'akli}}, \bibinfo {author} {\bibfnamefont
  {N.}~\bibnamefont {Pradhan}}, \ and\ \bibinfo {author} {\bibfnamefont
  {G.}~\bibnamefont {Iannacchione}},\ }\href {\doibase
  10.1103/PhysRevE.77.061701} {\bibfield  {journal} {\bibinfo  {journal} {Phys.
  Rev. E}\ }\textbf {\bibinfo {volume} {77}},\ \bibinfo {pages} {061701}
  (\bibinfo {year} {2008})}\BibitemShut {NoStop}%
\bibitem [{\citenamefont {Qazi}\ \emph {et~al.}(2010)\citenamefont {Qazi},
  \citenamefont {Karlsson},\ and\ \citenamefont {Rennie}}]{Qazi10}%
  \BibitemOpen
  \bibfield  {author} {\bibinfo {author} {\bibfnamefont {S.~J.~S.}\
  \bibnamefont {Qazi}}, \bibinfo {author} {\bibfnamefont {G.}~\bibnamefont
  {Karlsson}}, \ and\ \bibinfo {author} {\bibfnamefont {A.~R.}\ \bibnamefont
  {Rennie}},\ }\href {\doibase https://doi.org/10.1016/j.jcis.2010.04.033}
  {\bibfield  {journal} {\bibinfo  {journal} {Journal of Colloid and Interface
  Science}\ }\textbf {\bibinfo {volume} {348}},\ \bibinfo {pages} {80 }
  (\bibinfo {year} {2010})}\BibitemShut {NoStop}%
\bibitem [{\citenamefont {Aoshima}\ \emph {et~al.}(2012)\citenamefont
  {Aoshima}, \citenamefont {Ozaki},\ and\ \citenamefont {Satoh}}]{Aoshima12}%
  \BibitemOpen
  \bibfield  {author} {\bibinfo {author} {\bibfnamefont {M.}~\bibnamefont
  {Aoshima}}, \bibinfo {author} {\bibfnamefont {M.}~\bibnamefont {Ozaki}}, \
  and\ \bibinfo {author} {\bibfnamefont {A.}~\bibnamefont {Satoh}},\ }\href
  {\doibase 10.1021/jp301645x} {\bibfield  {journal} {\bibinfo  {journal} {The
  Journal of Physical Chemistry C}\ }\textbf {\bibinfo {volume} {116}},\
  \bibinfo {pages} {17862} (\bibinfo {year} {2012})},\ \Eprint
  {http://arxiv.org/abs/http://dx.doi.org/10.1021/jp301645x}
  {http://dx.doi.org/10.1021/jp301645x} \BibitemShut {NoStop}%
\bibitem [{\citenamefont {Lekkerkerker}\ and\ \citenamefont
  {Vroege}(2013)}]{Lekkerkerker12}%
  \BibitemOpen
  \bibfield  {author} {\bibinfo {author} {\bibfnamefont {H.~N.~W.}\
  \bibnamefont {Lekkerkerker}}\ and\ \bibinfo {author} {\bibfnamefont {G.~J.}\
  \bibnamefont {Vroege}},\ }\href {\doibase 10.1098/rsta.2012.0263} {\ \textbf
  {\bibinfo {volume} {371}} (\bibinfo {year} {2013}),\
  10.1098/rsta.2012.0263}\BibitemShut {NoStop}%
\bibitem [{\citenamefont {Blaak}\ \emph {et~al.}(2004)\citenamefont {Blaak},
  \citenamefont {Mulder},\ and\ \citenamefont {Frenkel}}]{Blaak04}%
  \BibitemOpen
  \bibfield  {author} {\bibinfo {author} {\bibfnamefont {R.}~\bibnamefont
  {Blaak}}, \bibinfo {author} {\bibfnamefont {B.~M.}\ \bibnamefont {Mulder}}, \
  and\ \bibinfo {author} {\bibfnamefont {D.}~\bibnamefont {Frenkel}},\
  }\href@noop {} {\bibfield  {journal} {\bibinfo  {journal} {The Journal of
  chemical physics}\ }\textbf {\bibinfo {volume} {120}},\ \bibinfo {pages}
  {5486} (\bibinfo {year} {2004})}\BibitemShut {NoStop}%
\bibitem [{\citenamefont {John}\ \emph {et~al.}(2004)\citenamefont {John},
  \citenamefont {Stroock},\ and\ \citenamefont {Escobedo}}]{John04}%
  \BibitemOpen
  \bibfield  {author} {\bibinfo {author} {\bibfnamefont {B.~S.}\ \bibnamefont
  {John}}, \bibinfo {author} {\bibfnamefont {A.}~\bibnamefont {Stroock}}, \
  and\ \bibinfo {author} {\bibfnamefont {F.~A.}\ \bibnamefont {Escobedo}},\
  }\href@noop {} {\bibfield  {journal} {\bibinfo  {journal} {The Journal of
  chemical physics}\ }\textbf {\bibinfo {volume} {120}},\ \bibinfo {pages}
  {9383} (\bibinfo {year} {2004})}\BibitemShut {NoStop}%
\bibitem [{\citenamefont {John}\ \emph {et~al.}(2008)\citenamefont {John},
  \citenamefont {Juhlin},\ and\ \citenamefont {Escobedo}}]{John08}%
  \BibitemOpen
  \bibfield  {author} {\bibinfo {author} {\bibfnamefont {B.~S.}\ \bibnamefont
  {John}}, \bibinfo {author} {\bibfnamefont {C.}~\bibnamefont {Juhlin}}, \ and\
  \bibinfo {author} {\bibfnamefont {F.~A.}\ \bibnamefont {Escobedo}},\
  }\href@noop {} {\bibfield  {journal} {\bibinfo  {journal} {The Journal of
  chemical physics}\ }\textbf {\bibinfo {volume} {128}},\ \bibinfo {pages}
  {044909} (\bibinfo {year} {2008})}\BibitemShut {NoStop}%
\bibitem [{\citenamefont {Duncan}\ \emph {et~al.}(2009)\citenamefont {Duncan},
  \citenamefont {Dennison}, \citenamefont {Masters},\ and\ \citenamefont
  {Wilson}}]{Duncan09}%
  \BibitemOpen
  \bibfield  {author} {\bibinfo {author} {\bibfnamefont {P.~D.}\ \bibnamefont
  {Duncan}}, \bibinfo {author} {\bibfnamefont {M.}~\bibnamefont {Dennison}},
  \bibinfo {author} {\bibfnamefont {A.~J.}\ \bibnamefont {Masters}}, \ and\
  \bibinfo {author} {\bibfnamefont {M.~R.}\ \bibnamefont {Wilson}},\ }\href
  {\doibase 10.1103/PhysRevE.79.031702} {\bibfield  {journal} {\bibinfo
  {journal} {Phys. Rev. E}\ }\textbf {\bibinfo {volume} {79}},\ \bibinfo
  {pages} {031702} (\bibinfo {year} {2009})}\BibitemShut {NoStop}%
\bibitem [{\citenamefont {Duncan}\ \emph {et~al.}(2011)\citenamefont {Duncan},
  \citenamefont {Masters},\ and\ \citenamefont {Wilson}}]{Duncan11}%
  \BibitemOpen
  \bibfield  {author} {\bibinfo {author} {\bibfnamefont {P.~D.}\ \bibnamefont
  {Duncan}}, \bibinfo {author} {\bibfnamefont {A.~J.}\ \bibnamefont {Masters}},
  \ and\ \bibinfo {author} {\bibfnamefont {M.~R.}\ \bibnamefont {Wilson}},\
  }\href {\doibase 10.1103/PhysRevE.84.011702} {\bibfield  {journal} {\bibinfo
  {journal} {Phys. Rev. E}\ }\textbf {\bibinfo {volume} {84}},\ \bibinfo
  {pages} {011702} (\bibinfo {year} {2011})}\BibitemShut {NoStop}%
\bibitem [{\citenamefont {Marechal}\ \emph {et~al.}(2012)\citenamefont
  {Marechal}, \citenamefont {Patti}, \citenamefont {Dennison},\ and\
  \citenamefont {Dijkstra}}]{Marechal12}%
  \BibitemOpen
  \bibfield  {author} {\bibinfo {author} {\bibfnamefont {M.}~\bibnamefont
  {Marechal}}, \bibinfo {author} {\bibfnamefont {A.}~\bibnamefont {Patti}},
  \bibinfo {author} {\bibfnamefont {M.}~\bibnamefont {Dennison}}, \ and\
  \bibinfo {author} {\bibfnamefont {M.}~\bibnamefont {Dijkstra}},\ }\href
  {\doibase 10.1103/PhysRevLett.108.206101} {\bibfield  {journal} {\bibinfo
  {journal} {Phys. Rev. Lett.}\ }\textbf {\bibinfo {volume} {108}},\ \bibinfo
  {pages} {206101} (\bibinfo {year} {2012})}\BibitemShut {NoStop}%
\bibitem [{\citenamefont {Wilson}\ \emph {et~al.}(2012)\citenamefont {Wilson},
  \citenamefont {Duncan}, \citenamefont {Dennison},\ and\ \citenamefont
  {Masters}}]{Wilson12}%
  \BibitemOpen
  \bibfield  {author} {\bibinfo {author} {\bibfnamefont {M.~R.}\ \bibnamefont
  {Wilson}}, \bibinfo {author} {\bibfnamefont {P.~D.}\ \bibnamefont {Duncan}},
  \bibinfo {author} {\bibfnamefont {M.}~\bibnamefont {Dennison}}, \ and\
  \bibinfo {author} {\bibfnamefont {A.~J.}\ \bibnamefont {Masters}},\ }\href
  {\doibase 10.1039/C2SM06962H} {\bibfield  {journal} {\bibinfo  {journal}
  {Soft Matter}\ }\textbf {\bibinfo {volume} {8}},\ \bibinfo {pages} {3348}
  (\bibinfo {year} {2012})}\BibitemShut {NoStop}%
\bibitem [{\citenamefont {Damasceno}\ \emph {et~al.}(2012)\citenamefont
  {Damasceno}, \citenamefont {Engel},\ and\ \citenamefont
  {Glotzer}}]{Glotzer12}%
  \BibitemOpen
  \bibfield  {author} {\bibinfo {author} {\bibfnamefont {P.~F.}\ \bibnamefont
  {Damasceno}}, \bibinfo {author} {\bibfnamefont {M.}~\bibnamefont {Engel}}, \
  and\ \bibinfo {author} {\bibfnamefont {S.~C.}\ \bibnamefont {Glotzer}},\
  }\href {\doibase 10.1126/science.1220869} {\bibfield  {journal} {\bibinfo
  {journal} {Science}\ }\textbf {\bibinfo {volume} {337}},\ \bibinfo {pages}
  {453} (\bibinfo {year} {2012})}\BibitemShut {NoStop}%
\bibitem [{\citenamefont {Cadotte}\ \emph {et~al.}(2016)\citenamefont
  {Cadotte}, \citenamefont {Dshemuchadse}, \citenamefont {Damasceno},
  \citenamefont {Newman},\ and\ \citenamefont {Glotzer}}]{Glotzer16a}%
  \BibitemOpen
  \bibfield  {author} {\bibinfo {author} {\bibfnamefont {A.~T.}\ \bibnamefont
  {Cadotte}}, \bibinfo {author} {\bibfnamefont {J.}~\bibnamefont
  {Dshemuchadse}}, \bibinfo {author} {\bibfnamefont {P.~F.}\ \bibnamefont
  {Damasceno}}, \bibinfo {author} {\bibfnamefont {R.~S.}\ \bibnamefont
  {Newman}}, \ and\ \bibinfo {author} {\bibfnamefont {S.~C.}\ \bibnamefont
  {Glotzer}},\ }\href@noop {} {\bibfield  {journal} {\bibinfo  {journal} {Soft
  Matter}\ }\textbf {\bibinfo {volume} {12}},\ \bibinfo {pages} {7073}
  (\bibinfo {year} {2016})}\BibitemShut {NoStop}%
\bibitem [{\citenamefont {Karas}\ \emph {et~al.}(2016)\citenamefont {Karas},
  \citenamefont {Glaser},\ and\ \citenamefont {Glotzer}}]{Glotzer16b}%
  \BibitemOpen
  \bibfield  {author} {\bibinfo {author} {\bibfnamefont {A.~S.}\ \bibnamefont
  {Karas}}, \bibinfo {author} {\bibfnamefont {J.}~\bibnamefont {Glaser}}, \
  and\ \bibinfo {author} {\bibfnamefont {S.~C.}\ \bibnamefont {Glotzer}},\
  }\href@noop {} {\bibfield  {journal} {\bibinfo  {journal} {Soft matter}\
  }\textbf {\bibinfo {volume} {12}},\ \bibinfo {pages} {5199} (\bibinfo {year}
  {2016})}\BibitemShut {NoStop}%
\bibitem [{\citenamefont {Fel}(1995)}]{Fel95}%
  \BibitemOpen
  \bibfield  {author} {\bibinfo {author} {\bibfnamefont {L.~G.}\ \bibnamefont
  {Fel}},\ }\href {\doibase 10.1103/PhysRevE.52.702} {\bibfield  {journal}
  {\bibinfo  {journal} {Phys. Rev. E}\ }\textbf {\bibinfo {volume} {52}},\
  \bibinfo {pages} {702} (\bibinfo {year} {1995})}\BibitemShut {NoStop}%
\bibitem [{\citenamefont {Michel}\ and\ \citenamefont
  {Zhilinski{\i}}(2001)}]{Michel01}%
  \BibitemOpen
  \bibfield  {author} {\bibinfo {author} {\bibfnamefont {L.}~\bibnamefont
  {Michel}}\ and\ \bibinfo {author} {\bibfnamefont {B.}~\bibnamefont
  {Zhilinski{\i}}},\ }\href@noop {} {\bibfield  {journal} {\bibinfo  {journal}
  {Physics Reports}\ }\textbf {\bibinfo {volume} {341}},\ \bibinfo {pages} {11}
  (\bibinfo {year} {2001})}\BibitemShut {NoStop}%
\bibitem [{\citenamefont {Haji-Akbari}\ and\ \citenamefont
  {Glotzer}(2015)}]{HajiAkbariGlotzer15}%
  \BibitemOpen
  \bibfield  {author} {\bibinfo {author} {\bibfnamefont {A.}~\bibnamefont
  {Haji-Akbari}}\ and\ \bibinfo {author} {\bibfnamefont {S.~C.}\ \bibnamefont
  {Glotzer}},\ }\href {http://stacks.iop.org/1751-8121/48/i=48/a=485201}
  {\bibfield  {journal} {\bibinfo  {journal} {Journal of Physics A:
  Mathematical and Theoretical}\ }\textbf {\bibinfo {volume} {48}},\ \bibinfo
  {pages} {485201} (\bibinfo {year} {2015})}\BibitemShut {NoStop}%
\bibitem [{\citenamefont {Mermin}(1979)}]{Mermin79}%
  \BibitemOpen
  \bibfield  {author} {\bibinfo {author} {\bibfnamefont {N.~D.}\ \bibnamefont
  {Mermin}},\ }\href {\doibase 10.1103/RevModPhys.51.591} {\bibfield  {journal}
  {\bibinfo  {journal} {Rev. Mod. Phys.}\ }\textbf {\bibinfo {volume} {51}},\
  \bibinfo {pages} {591} (\bibinfo {year} {1979})}\BibitemShut {NoStop}%
\bibitem [{\citenamefont {Michel}(1980)}]{Michel80}%
  \BibitemOpen
  \bibfield  {author} {\bibinfo {author} {\bibfnamefont {L.}~\bibnamefont
  {Michel}},\ }\href {\doibase 10.1103/RevModPhys.52.617} {\bibfield  {journal}
  {\bibinfo  {journal} {Rev. Mod. Phys.}\ }\textbf {\bibinfo {volume} {52}},\
  \bibinfo {pages} {617} (\bibinfo {year} {1980})}\BibitemShut {NoStop}%
\bibitem [{\citenamefont {Nelson}\ and\ \citenamefont
  {Toner}(1981)}]{NelsonToner81}%
  \BibitemOpen
  \bibfield  {author} {\bibinfo {author} {\bibfnamefont {D.~R.}\ \bibnamefont
  {Nelson}}\ and\ \bibinfo {author} {\bibfnamefont {J.}~\bibnamefont {Toner}},\
  }\href@noop {} {\bibfield  {journal} {\bibinfo  {journal} {Physical Review
  B}\ }\textbf {\bibinfo {volume} {24}},\ \bibinfo {pages} {363} (\bibinfo
  {year} {1981})}\BibitemShut {NoStop}%
\bibitem [{\citenamefont {Steinhardt}\ \emph {et~al.}(1981)\citenamefont
  {Steinhardt}, \citenamefont {Nelson},\ and\ \citenamefont
  {Ronchetti}}]{SteinhardtNelson1981}%
  \BibitemOpen
  \bibfield  {author} {\bibinfo {author} {\bibfnamefont {P.~J.}\ \bibnamefont
  {Steinhardt}}, \bibinfo {author} {\bibfnamefont {D.~R.}\ \bibnamefont
  {Nelson}}, \ and\ \bibinfo {author} {\bibfnamefont {M.}~\bibnamefont
  {Ronchetti}},\ }\href {\doibase 10.1103/PhysRevLett.47.1297} {\bibfield
  {journal} {\bibinfo  {journal} {Phys. Rev. Lett.}\ }\textbf {\bibinfo
  {volume} {47}},\ \bibinfo {pages} {1297} (\bibinfo {year}
  {1981})}\BibitemShut {NoStop}%
\bibitem [{\citenamefont {Romano}(2006)}]{Romano06}%
  \BibitemOpen
  \bibfield  {author} {\bibinfo {author} {\bibfnamefont {S.}~\bibnamefont
  {Romano}},\ }\href@noop {} {\bibfield  {journal} {\bibinfo  {journal}
  {Physical Review E}\ }\textbf {\bibinfo {volume} {74}},\ \bibinfo {pages}
  {011704} (\bibinfo {year} {2006})}\BibitemShut {NoStop}%
\bibitem [{\citenamefont {Romano}(2008)}]{Romano08}%
  \BibitemOpen
  \bibfield  {author} {\bibinfo {author} {\bibfnamefont {S.}~\bibnamefont
  {Romano}},\ }\href@noop {} {\bibfield  {journal} {\bibinfo  {journal}
  {Physical Review E}\ }\textbf {\bibinfo {volume} {77}},\ \bibinfo {pages}
  {021704} (\bibinfo {year} {2008})}\BibitemShut {NoStop}%
\bibitem [{\citenamefont {Trojanowski}\ \emph {et~al.}(2012)\citenamefont
  {Trojanowski}, \citenamefont {Paj\c{a}k}, \citenamefont {Longa},\ and\
  \citenamefont {Wydro}}]{TrojanowskiLonga12}%
  \BibitemOpen
  \bibfield  {author} {\bibinfo {author} {\bibfnamefont {K.}~\bibnamefont
  {Trojanowski}}, \bibinfo {author} {\bibfnamefont {G.}~\bibnamefont
  {Paj\c{a}k}}, \bibinfo {author} {\bibfnamefont {L.}~\bibnamefont {Longa}}, \
  and\ \bibinfo {author} {\bibfnamefont {T.}~\bibnamefont {Wydro}},\ }\href
  {\doibase 10.1103/PhysRevE.86.011704} {\bibfield  {journal} {\bibinfo
  {journal} {Phys. Rev. E}\ }\textbf {\bibinfo {volume} {86}},\ \bibinfo
  {pages} {011704} (\bibinfo {year} {2012})}\BibitemShut {NoStop}%
\bibitem [{\citenamefont {Stallinga}\ and\ \citenamefont
  {Vertogen}(1994)}]{Stallinga94}%
  \BibitemOpen
  \bibfield  {author} {\bibinfo {author} {\bibfnamefont {S.}~\bibnamefont
  {Stallinga}}\ and\ \bibinfo {author} {\bibfnamefont {G.}~\bibnamefont
  {Vertogen}},\ }\href {\doibase 10.1103/PhysRevE.49.1483} {\bibfield
  {journal} {\bibinfo  {journal} {Phys. Rev. E}\ }\textbf {\bibinfo {volume}
  {49}},\ \bibinfo {pages} {1483} (\bibinfo {year} {1994})}\BibitemShut
  {NoStop}%
\bibitem [{\citenamefont {Brand}\ \emph {et~al.}(2005)\citenamefont {Brand},
  \citenamefont {Cladis},\ and\ \citenamefont {Pleiner}}]{Brand05}%
  \BibitemOpen
  \bibfield  {author} {\bibinfo {author} {\bibfnamefont {H.~R.}\ \bibnamefont
  {Brand}}, \bibinfo {author} {\bibfnamefont {P.~E.}\ \bibnamefont {Cladis}}, \
  and\ \bibinfo {author} {\bibfnamefont {H.}~\bibnamefont {Pleiner}},\ }\href
  {\doibase 10.1080/00150190490509656} {\bibfield  {journal} {\bibinfo
  {journal} {Ferroelectrics}\ }\textbf {\bibinfo {volume} {315}},\ \bibinfo
  {pages} {165} (\bibinfo {year} {2005})},\ \Eprint
  {http://arxiv.org/abs/http://dx.doi.org/10.1080/00150190490509656}
  {http://dx.doi.org/10.1080/00150190490509656} \BibitemShut {NoStop}%
\bibitem [{\citenamefont {Liu}\ \emph {et~al.}(2016)\citenamefont {Liu},
  \citenamefont {Nissinen}, \citenamefont {Slager}, \citenamefont {Wu},\ and\
  \citenamefont {Zaanen}}]{Liu16}%
  \BibitemOpen
  \bibfield  {author} {\bibinfo {author} {\bibfnamefont {K.}~\bibnamefont
  {Liu}}, \bibinfo {author} {\bibfnamefont {J.}~\bibnamefont {Nissinen}},
  \bibinfo {author} {\bibfnamefont {R.-J.}\ \bibnamefont {Slager}}, \bibinfo
  {author} {\bibfnamefont {K.}~\bibnamefont {Wu}}, \ and\ \bibinfo {author}
  {\bibfnamefont {J.}~\bibnamefont {Zaanen}},\ }\href {\doibase
  10.1103/PhysRevX.6.041025} {\bibfield  {journal} {\bibinfo  {journal} {Phys.
  Rev. X}\ }\textbf {\bibinfo {volume} {6}},\ \bibinfo {pages} {041025}
  (\bibinfo {year} {2016})}\BibitemShut {NoStop}%
\bibitem [{\citenamefont {Nissinen}\ \emph {et~al.}(2016)\citenamefont
  {Nissinen}, \citenamefont {Liu}, \citenamefont {Slager}, \citenamefont {Wu},\
  and\ \citenamefont {Zaanen}}]{Nissinen16}%
  \BibitemOpen
  \bibfield  {author} {\bibinfo {author} {\bibfnamefont {J.}~\bibnamefont
  {Nissinen}}, \bibinfo {author} {\bibfnamefont {K.}~\bibnamefont {Liu}},
  \bibinfo {author} {\bibfnamefont {R.-J.}\ \bibnamefont {Slager}}, \bibinfo
  {author} {\bibfnamefont {K.}~\bibnamefont {Wu}}, \ and\ \bibinfo {author}
  {\bibfnamefont {J.}~\bibnamefont {Zaanen}},\ }\href {\doibase
  10.1103/PhysRevE.94.022701} {\bibfield  {journal} {\bibinfo  {journal} {Phys.
  Rev. E}\ }\textbf {\bibinfo {volume} {94}},\ \bibinfo {pages} {022701}
  (\bibinfo {year} {2016})}\BibitemShut {NoStop}%
\bibitem [{\citenamefont {Kogut}(1979)}]{Kogut79}%
  \BibitemOpen
  \bibfield  {author} {\bibinfo {author} {\bibfnamefont {J.~B.}\ \bibnamefont
  {Kogut}},\ }\href {\doibase 10.1103/RevModPhys.51.659} {\bibfield  {journal}
  {\bibinfo  {journal} {Rev. Mod. Phys.}\ }\textbf {\bibinfo {volume} {51}},\
  \bibinfo {pages} {659} (\bibinfo {year} {1979})}\BibitemShut {NoStop}%
\bibitem [{\citenamefont {Lammert}\ \emph {et~al.}(1993)\citenamefont
  {Lammert}, \citenamefont {Rokhsar},\ and\ \citenamefont
  {Toner}}]{LammertRoksharToner93}%
  \BibitemOpen
  \bibfield  {author} {\bibinfo {author} {\bibfnamefont {P.~E.}\ \bibnamefont
  {Lammert}}, \bibinfo {author} {\bibfnamefont {D.~S.}\ \bibnamefont
  {Rokhsar}}, \ and\ \bibinfo {author} {\bibfnamefont {J.}~\bibnamefont
  {Toner}},\ }\href {\doibase 10.1103/PhysRevLett.70.1650} {\bibfield
  {journal} {\bibinfo  {journal} {Phys. Rev. Lett.}\ }\textbf {\bibinfo
  {volume} {70}},\ \bibinfo {pages} {1650} (\bibinfo {year}
  {1993})}\BibitemShut {NoStop}%
\bibitem [{\citenamefont {Lammert}\ \emph {et~al.}(1995)\citenamefont
  {Lammert}, \citenamefont {Rokhsar},\ and\ \citenamefont
  {Toner}}]{LammertRoksharToner95}%
  \BibitemOpen
  \bibfield  {author} {\bibinfo {author} {\bibfnamefont {P.~E.}\ \bibnamefont
  {Lammert}}, \bibinfo {author} {\bibfnamefont {D.~S.}\ \bibnamefont
  {Rokhsar}}, \ and\ \bibinfo {author} {\bibfnamefont {J.}~\bibnamefont
  {Toner}},\ }\href {\doibase 10.1103/PhysRevE.52.1778} {\bibfield  {journal}
  {\bibinfo  {journal} {Phys. Rev. E}\ }\textbf {\bibinfo {volume} {52}},\
  \bibinfo {pages} {1778} (\bibinfo {year} {1995})}\BibitemShut {NoStop}%
\bibitem [{\citenamefont {Toner}\ \emph {et~al.}(1995)\citenamefont {Toner},
  \citenamefont {Lammert},\ and\ \citenamefont
  {Rokhsar}}]{TonerLammertRokshar95b}%
  \BibitemOpen
  \bibfield  {author} {\bibinfo {author} {\bibfnamefont {J.}~\bibnamefont
  {Toner}}, \bibinfo {author} {\bibfnamefont {P.~E.}\ \bibnamefont {Lammert}},
  \ and\ \bibinfo {author} {\bibfnamefont {D.~S.}\ \bibnamefont {Rokhsar}},\
  }\href {\doibase 10.1103/PhysRevE.52.1801} {\bibfield  {journal} {\bibinfo
  {journal} {Phys. Rev. E}\ }\textbf {\bibinfo {volume} {52}},\ \bibinfo
  {pages} {1801} (\bibinfo {year} {1995})}\BibitemShut {NoStop}%
\bibitem [{\citenamefont {Senthil}\ and\ \citenamefont
  {Fisher}(2000)}]{SenthilFisher00}%
  \BibitemOpen
  \bibfield  {author} {\bibinfo {author} {\bibfnamefont {T.}~\bibnamefont
  {Senthil}}\ and\ \bibinfo {author} {\bibfnamefont {M.~P.~A.}\ \bibnamefont
  {Fisher}},\ }\href {\doibase 10.1103/PhysRevB.62.7850} {\bibfield  {journal}
  {\bibinfo  {journal} {Phys. Rev. B}\ }\textbf {\bibinfo {volume} {62}},\
  \bibinfo {pages} {7850} (\bibinfo {year} {2000})}\BibitemShut {NoStop}%
\bibitem [{\citenamefont {Senthil}\ and\ \citenamefont
  {Fisher}(2001)}]{SenthilFisher01}%
  \BibitemOpen
  \bibfield  {author} {\bibinfo {author} {\bibfnamefont {T.}~\bibnamefont
  {Senthil}}\ and\ \bibinfo {author} {\bibfnamefont {M.~P.~A.}\ \bibnamefont
  {Fisher}},\ }\href {\doibase 10.1103/PhysRevB.63.134521} {\bibfield
  {journal} {\bibinfo  {journal} {Phys. Rev. B}\ }\textbf {\bibinfo {volume}
  {63}},\ \bibinfo {pages} {134521} (\bibinfo {year} {2001})}\BibitemShut
  {NoStop}%
\bibitem [{\citenamefont {Demler}\ \emph {et~al.}(2002)\citenamefont {Demler},
  \citenamefont {Nayak}, \citenamefont {Kee}, \citenamefont {Kim},\ and\
  \citenamefont {Senthil}}]{DemlerSenthil05}%
  \BibitemOpen
  \bibfield  {author} {\bibinfo {author} {\bibfnamefont {E.}~\bibnamefont
  {Demler}}, \bibinfo {author} {\bibfnamefont {C.}~\bibnamefont {Nayak}},
  \bibinfo {author} {\bibfnamefont {H.-Y.}\ \bibnamefont {Kee}}, \bibinfo
  {author} {\bibfnamefont {Y.~B.}\ \bibnamefont {Kim}}, \ and\ \bibinfo
  {author} {\bibfnamefont {T.}~\bibnamefont {Senthil}},\ }\href {\doibase
  10.1103/PhysRevB.65.155103} {\bibfield  {journal} {\bibinfo  {journal} {Phys.
  Rev. B}\ }\textbf {\bibinfo {volume} {65}},\ \bibinfo {pages} {155103}
  (\bibinfo {year} {2002})}\BibitemShut {NoStop}%
\bibitem [{\citenamefont {{Z. Nussinov}}\ and\ \citenamefont {{J.
  Zaanen}}(2002)}]{ZaanenNussinov02}%
  \BibitemOpen
  \bibfield  {author} {\bibinfo {author} {\bibnamefont {{Z. Nussinov}}}\ and\
  \bibinfo {author} {\bibnamefont {{J. Zaanen}}},\ }\href {\doibase
  10.1051/jp4:20020405} {\bibfield  {journal} {\bibinfo  {journal} {J. Phys. IV
  France}\ }\textbf {\bibinfo {volume} {12}},\ \bibinfo {pages} {245} (\bibinfo
  {year} {2002})}\BibitemShut {NoStop}%
\bibitem [{\citenamefont {Podolsky}\ and\ \citenamefont
  {Demler}(2005)}]{PodolskyDemler05}%
  \BibitemOpen
  \bibfield  {author} {\bibinfo {author} {\bibfnamefont {D.}~\bibnamefont
  {Podolsky}}\ and\ \bibinfo {author} {\bibfnamefont {E.}~\bibnamefont
  {Demler}},\ }\href@noop {} {\bibfield  {journal} {\bibinfo  {journal} {New
  Journal of Physics}\ }\textbf {\bibinfo {volume} {7}},\ \bibinfo {pages} {59}
  (\bibinfo {year} {2005})}\BibitemShut {NoStop}%
\bibitem [{\citenamefont {Liu}\ \emph {et~al.}(2015)\citenamefont {Liu},
  \citenamefont {Nissinen}, \citenamefont {Nussinov}, \citenamefont {Slager},
  \citenamefont {Wu},\ and\ \citenamefont {Zaanen}}]{Liu15}%
  \BibitemOpen
  \bibfield  {author} {\bibinfo {author} {\bibfnamefont {K.}~\bibnamefont
  {Liu}}, \bibinfo {author} {\bibfnamefont {J.}~\bibnamefont {Nissinen}},
  \bibinfo {author} {\bibfnamefont {Z.}~\bibnamefont {Nussinov}}, \bibinfo
  {author} {\bibfnamefont {R.-J.}\ \bibnamefont {Slager}}, \bibinfo {author}
  {\bibfnamefont {K.}~\bibnamefont {Wu}}, \ and\ \bibinfo {author}
  {\bibfnamefont {J.}~\bibnamefont {Zaanen}},\ }\href {\doibase
  10.1103/PhysRevB.91.075103} {\bibfield  {journal} {\bibinfo  {journal} {Phys.
  Rev. B}\ }\textbf {\bibinfo {volume} {91}},\ \bibinfo {pages} {075103}
  (\bibinfo {year} {2015})}\BibitemShut {NoStop}%
\bibitem [{\citenamefont {Beekman}\ \emph {et~al.}(2017)\citenamefont
  {Beekman}, \citenamefont {Nissinen}, \citenamefont {Wu}, \citenamefont {Liu},
  \citenamefont {Slager}, \citenamefont {Nussinov}, \citenamefont {Cvetkovic},\
  and\ \citenamefont {Zaanen}}]{Beekman17}%
  \BibitemOpen
  \bibfield  {author} {\bibinfo {author} {\bibfnamefont {A.}~\bibnamefont
  {Beekman}}, \bibinfo {author} {\bibfnamefont {J.}~\bibnamefont {Nissinen}},
  \bibinfo {author} {\bibfnamefont {K.}~\bibnamefont {Wu}}, \bibinfo {author}
  {\bibfnamefont {K.}~\bibnamefont {Liu}}, \bibinfo {author} {\bibfnamefont
  {R.-J.}\ \bibnamefont {Slager}}, \bibinfo {author} {\bibfnamefont
  {Z.}~\bibnamefont {Nussinov}}, \bibinfo {author} {\bibfnamefont
  {V.}~\bibnamefont {Cvetkovic}}, \ and\ \bibinfo {author} {\bibfnamefont
  {J.}~\bibnamefont {Zaanen}},\ }\href {\doibase
  https://doi.org/10.1016/j.physrep.2017.03.004} {\bibfield  {journal}
  {\bibinfo  {journal} {Physics Reports}\ }\textbf {\bibinfo {volume} {683}},\
  \bibinfo {pages} {1 } (\bibinfo {year} {2017})}\BibitemShut {NoStop}%
\bibitem [{\citenamefont {Liu}\ \emph {et~al.}(2017)\citenamefont {Liu},
  \citenamefont {Nissinen}, \citenamefont {de~Boer}, \citenamefont {Slager},\
  and\ \citenamefont {Zaanen}}]{Liu17}%
  \BibitemOpen
  \bibfield  {author} {\bibinfo {author} {\bibfnamefont {K.}~\bibnamefont
  {Liu}}, \bibinfo {author} {\bibfnamefont {J.}~\bibnamefont {Nissinen}},
  \bibinfo {author} {\bibfnamefont {J.}~\bibnamefont {de~Boer}}, \bibinfo
  {author} {\bibfnamefont {R.-J.}\ \bibnamefont {Slager}}, \ and\ \bibinfo
  {author} {\bibfnamefont {J.}~\bibnamefont {Zaanen}},\ }\href {\doibase
  10.1103/PhysRevE.95.022704} {\bibfield  {journal} {\bibinfo  {journal} {Phys.
  Rev. E}\ }\textbf {\bibinfo {volume} {95}},\ \bibinfo {pages} {022704}
  (\bibinfo {year} {2017})}\BibitemShut {NoStop}%
\bibitem [{\citenamefont {Dressel}\ \emph {et~al.}(2014)\citenamefont
  {Dressel}, \citenamefont {Reppe}, \citenamefont {Prehm}, \citenamefont
  {Brautzsch},\ and\ \citenamefont {Tschierske}}]{Dressel14}%
  \BibitemOpen
  \bibfield  {author} {\bibinfo {author} {\bibfnamefont {C.}~\bibnamefont
  {Dressel}}, \bibinfo {author} {\bibfnamefont {T.}~\bibnamefont {Reppe}},
  \bibinfo {author} {\bibfnamefont {M.}~\bibnamefont {Prehm}}, \bibinfo
  {author} {\bibfnamefont {M.}~\bibnamefont {Brautzsch}}, \ and\ \bibinfo
  {author} {\bibfnamefont {C.}~\bibnamefont {Tschierske}},\ }\href@noop {}
  {\bibfield  {journal} {\bibinfo  {journal} {Nature chemistry}\ }\textbf
  {\bibinfo {volume} {6}},\ \bibinfo {pages} {971} (\bibinfo {year}
  {2014})}\BibitemShut {NoStop}%
\bibitem [{\citenamefont {Glotzer}\ and\ \citenamefont
  {Solomon}(2007)}]{Glotzer07}%
  \BibitemOpen
  \bibfield  {author} {\bibinfo {author} {\bibfnamefont {S.~C.}\ \bibnamefont
  {Glotzer}}\ and\ \bibinfo {author} {\bibfnamefont {M.~J.}\ \bibnamefont
  {Solomon}},\ }\href@noop {} {\bibfield  {journal} {\bibinfo  {journal}
  {Nature materials}\ }\textbf {\bibinfo {volume} {6}},\ \bibinfo {pages} {557}
  (\bibinfo {year} {2007})}\BibitemShut {NoStop}%
\bibitem [{\citenamefont {Glotzer}\ and\ \citenamefont
  {Anderson}(2010)}]{Glotzer10}%
  \BibitemOpen
  \bibfield  {author} {\bibinfo {author} {\bibfnamefont {S.~C.}\ \bibnamefont
  {Glotzer}}\ and\ \bibinfo {author} {\bibfnamefont {J.~A.}\ \bibnamefont
  {Anderson}},\ }\href@noop {} {\bibfield  {journal} {\bibinfo  {journal}
  {Nature materials}\ }\textbf {\bibinfo {volume} {9}},\ \bibinfo {pages} {885}
  (\bibinfo {year} {2010})}\BibitemShut {NoStop}%
\bibitem [{\citenamefont {Kraft}\ \emph {et~al.}(2009)\citenamefont {Kraft},
  \citenamefont {Groenewold},\ and\ \citenamefont {Kegel}}]{Kraft09}%
  \BibitemOpen
  \bibfield  {author} {\bibinfo {author} {\bibfnamefont {D.~J.}\ \bibnamefont
  {Kraft}}, \bibinfo {author} {\bibfnamefont {J.}~\bibnamefont {Groenewold}}, \
  and\ \bibinfo {author} {\bibfnamefont {W.~K.}\ \bibnamefont {Kegel}},\ }\href
  {\doibase 10.1039/B910593J} {\bibfield  {journal} {\bibinfo  {journal} {Soft
  Matter}\ }\textbf {\bibinfo {volume} {5}},\ \bibinfo {pages} {3823} (\bibinfo
  {year} {2009})}\BibitemShut {NoStop}%
\bibitem [{\citenamefont {Zwanikken}\ \emph {et~al.}(2011)\citenamefont
  {Zwanikken}, \citenamefont {Ioannidou}, \citenamefont {Kraft},\ and\
  \citenamefont {van Roij}}]{Kraft11}%
  \BibitemOpen
  \bibfield  {author} {\bibinfo {author} {\bibfnamefont {J.}~\bibnamefont
  {Zwanikken}}, \bibinfo {author} {\bibfnamefont {K.}~\bibnamefont
  {Ioannidou}}, \bibinfo {author} {\bibfnamefont {D.}~\bibnamefont {Kraft}}, \
  and\ \bibinfo {author} {\bibfnamefont {R.}~\bibnamefont {van Roij}},\ }\href
  {\doibase 10.1039/C1SM05779K} {\bibfield  {journal} {\bibinfo  {journal}
  {Soft Matter}\ }\textbf {\bibinfo {volume} {7}},\ \bibinfo {pages} {11093}
  (\bibinfo {year} {2011})}\BibitemShut {NoStop}%
\bibitem [{\citenamefont {Mark}\ \emph {et~al.}(2013)\citenamefont {Mark},
  \citenamefont {Gibbs}, \citenamefont {Lee},\ and\ \citenamefont
  {Fischer}}]{Mark13}%
  \BibitemOpen
  \bibfield  {author} {\bibinfo {author} {\bibfnamefont {A.~G.}\ \bibnamefont
  {Mark}}, \bibinfo {author} {\bibfnamefont {J.~G.}\ \bibnamefont {Gibbs}},
  \bibinfo {author} {\bibfnamefont {T.-C.}\ \bibnamefont {Lee}}, \ and\
  \bibinfo {author} {\bibfnamefont {P.}~\bibnamefont {Fischer}},\ }\href
  {http://dx.doi.org/10.1038/nmat3685} {\bibfield  {journal} {\bibinfo
  {journal} {Nat Mater}\ }\textbf {\bibinfo {volume} {12}},\ \bibinfo {pages}
  {802} (\bibinfo {year} {2013})}\BibitemShut {NoStop}%
\bibitem [{\citenamefont {Huang}\ \emph {et~al.}(2015)\citenamefont {Huang},
  \citenamefont {Hsu}, \citenamefont {Wang}, \citenamefont {Mei}, \citenamefont
  {Dong}, \citenamefont {Li}, \citenamefont {Li}, \citenamefont {Liu},
  \citenamefont {Zhang}, \citenamefont {Aida}, \citenamefont {Zhang},
  \citenamefont {Yue},\ and\ \citenamefont {Cheng}}]{Huang15}%
  \BibitemOpen
  \bibfield  {author} {\bibinfo {author} {\bibfnamefont {M.}~\bibnamefont
  {Huang}}, \bibinfo {author} {\bibfnamefont {C.-H.}\ \bibnamefont {Hsu}},
  \bibinfo {author} {\bibfnamefont {J.}~\bibnamefont {Wang}}, \bibinfo {author}
  {\bibfnamefont {S.}~\bibnamefont {Mei}}, \bibinfo {author} {\bibfnamefont
  {X.}~\bibnamefont {Dong}}, \bibinfo {author} {\bibfnamefont {Y.}~\bibnamefont
  {Li}}, \bibinfo {author} {\bibfnamefont {M.}~\bibnamefont {Li}}, \bibinfo
  {author} {\bibfnamefont {H.}~\bibnamefont {Liu}}, \bibinfo {author}
  {\bibfnamefont {W.}~\bibnamefont {Zhang}}, \bibinfo {author} {\bibfnamefont
  {T.}~\bibnamefont {Aida}}, \bibinfo {author} {\bibfnamefont {W.-B.}\
  \bibnamefont {Zhang}}, \bibinfo {author} {\bibfnamefont {K.}~\bibnamefont
  {Yue}}, \ and\ \bibinfo {author} {\bibfnamefont {S.~Z.~D.}\ \bibnamefont
  {Cheng}},\ }\href {\doibase 10.1126/science.aaa2421} {\bibfield  {journal}
  {\bibinfo  {journal} {Science}\ }\textbf {\bibinfo {volume} {348}},\ \bibinfo
  {pages} {424} (\bibinfo {year} {2015})}\BibitemShut {NoStop}%
\bibitem [{\citenamefont {Blaak}\ \emph {et~al.}(1999)\citenamefont {Blaak},
  \citenamefont {Frenkel},\ and\ \citenamefont {Mulder}}]{Blaak99}%
  \BibitemOpen
  \bibfield  {author} {\bibinfo {author} {\bibfnamefont {R.}~\bibnamefont
  {Blaak}}, \bibinfo {author} {\bibfnamefont {D.}~\bibnamefont {Frenkel}}, \
  and\ \bibinfo {author} {\bibfnamefont {B.~M.}\ \bibnamefont {Mulder}},\
  }\href {\doibase 10.1063/1.479104} {\bibfield  {journal} {\bibinfo  {journal}
  {The Journal of Chemical Physics}\ }\textbf {\bibinfo {volume} {110}},\
  \bibinfo {pages} {11652} (\bibinfo {year} {1999})},\ \Eprint
  {http://arxiv.org/abs/http://dx.doi.org/10.1063/1.479104}
  {http://dx.doi.org/10.1063/1.479104} \BibitemShut {NoStop}%
\bibitem [{\citenamefont {Romano}(2016)}]{Romano16}%
  \BibitemOpen
  \bibfield  {author} {\bibinfo {author} {\bibfnamefont {S.}~\bibnamefont
  {Romano}},\ }\href {\doibase 10.1103/PhysRevE.94.042702} {\bibfield
  {journal} {\bibinfo  {journal} {Phys. Rev. E}\ }\textbf {\bibinfo {volume}
  {94}},\ \bibinfo {pages} {042702} (\bibinfo {year} {2016})}\BibitemShut
  {NoStop}%
\bibitem [{\citenamefont {Mettout}(2006)}]{Mettout06}%
  \BibitemOpen
  \bibfield  {author} {\bibinfo {author} {\bibfnamefont {B.}~\bibnamefont
  {Mettout}},\ }\href {\doibase 10.1103/PhysRevE.74.041701} {\bibfield
  {journal} {\bibinfo  {journal} {Phys. Rev. E}\ }\textbf {\bibinfo {volume}
  {74}},\ \bibinfo {pages} {041701} (\bibinfo {year} {2006})}\BibitemShut
  {NoStop}%
\bibitem [{\citenamefont {Elitzur}(1975)}]{Elitzur75}%
  \BibitemOpen
  \bibfield  {author} {\bibinfo {author} {\bibfnamefont {S.}~\bibnamefont
  {Elitzur}},\ }\href {\doibase 10.1103/PhysRevD.12.3978} {\bibfield  {journal}
  {\bibinfo  {journal} {Phys. Rev. D}\ }\textbf {\bibinfo {volume} {12}},\
  \bibinfo {pages} {3978} (\bibinfo {year} {1975})}\BibitemShut {NoStop}%
\bibitem [{\citenamefont {Fradkin}\ and\ \citenamefont
  {Shenker}(1979)}]{FradkinShenker79}%
  \BibitemOpen
  \bibfield  {author} {\bibinfo {author} {\bibfnamefont {E.}~\bibnamefont
  {Fradkin}}\ and\ \bibinfo {author} {\bibfnamefont {S.~H.}\ \bibnamefont
  {Shenker}},\ }\href {\doibase 10.1103/PhysRevD.19.3682} {\bibfield  {journal}
  {\bibinfo  {journal} {Phys. Rev. D}\ }\textbf {\bibinfo {volume} {19}},\
  \bibinfo {pages} {3682} (\bibinfo {year} {1979})}\BibitemShut {NoStop}%
\bibitem [{\citenamefont {Kleman}\ and\ \citenamefont
  {Friedel}(2008)}]{Kleman08}%
  \BibitemOpen
  \bibfield  {author} {\bibinfo {author} {\bibfnamefont {M.}~\bibnamefont
  {Kleman}}\ and\ \bibinfo {author} {\bibfnamefont {J.}~\bibnamefont
  {Friedel}},\ }\href {\doibase 10.1103/RevModPhys.80.61} {\bibfield  {journal}
  {\bibinfo  {journal} {Rev. Mod. Phys.}\ }\textbf {\bibinfo {volume} {80}},\
  \bibinfo {pages} {61} (\bibinfo {year} {2008})}\BibitemShut {NoStop}%
\bibitem [{\citenamefont {Butler}(2012)}]{BookButler12}%
  \BibitemOpen
  \bibfield  {author} {\bibinfo {author} {\bibfnamefont {P.}~\bibnamefont
  {Butler}},\ }\href {https://books.google.nl/books?id=jmDSBwAAQBAJ} {\emph
  {\bibinfo {title} {Point Group Symmetry Applications: Methods and Tables}}}\
  (\bibinfo  {publisher} {Springer US},\ \bibinfo {year} {2012})\BibitemShut
  {NoStop}%
\bibitem [{\citenamefont {Bishop}(1993)}]{BookBishop93}%
  \BibitemOpen
  \bibfield  {author} {\bibinfo {author} {\bibfnamefont {D.}~\bibnamefont
  {Bishop}},\ }\href {https://books.google.nl/books?id=pn\_1DAAAQBAJ} {\emph
  {\bibinfo {title} {Group Theory and Chemistry}}},\ Dover Books on Chemistry
  Series\ (\bibinfo  {publisher} {Dover Publications},\ \bibinfo {year}
  {1993})\BibitemShut {NoStop}%
\bibitem [{\citenamefont {Ferrenberg}\ and\ \citenamefont
  {Swendsen}(1988)}]{Ferrenberg88}%
  \BibitemOpen
  \bibfield  {author} {\bibinfo {author} {\bibfnamefont {A.~M.}\ \bibnamefont
  {Ferrenberg}}\ and\ \bibinfo {author} {\bibfnamefont {R.~H.}\ \bibnamefont
  {Swendsen}},\ }\href {\doibase 10.1103/PhysRevLett.61.2635} {\bibfield
  {journal} {\bibinfo  {journal} {Phys. Rev. Lett.}\ }\textbf {\bibinfo
  {volume} {61}},\ \bibinfo {pages} {2635} (\bibinfo {year}
  {1988})}\BibitemShut {NoStop}%
\bibitem [{\citenamefont {Newman}\ and\ \citenamefont
  {Barkema}(1999)}]{BookNewman99}%
  \BibitemOpen
  \bibfield  {author} {\bibinfo {author} {\bibfnamefont {M.}~\bibnamefont
  {Newman}}\ and\ \bibinfo {author} {\bibfnamefont {G.}~\bibnamefont
  {Barkema}},\ }\href {https://books.google.de/books?id=J5aLdDN4uFwC} {\emph
  {\bibinfo {title} {Monte Carlo Methods in Statistical Physics}}}\ (\bibinfo
  {publisher} {Clarendon Press},\ \bibinfo {year} {1999})\BibitemShut {NoStop}%
\bibitem [{\citenamefont {Lee}\ and\ \citenamefont {Kosterlitz}(1990)}]{Lee90}%
  \BibitemOpen
  \bibfield  {author} {\bibinfo {author} {\bibfnamefont {J.}~\bibnamefont
  {Lee}}\ and\ \bibinfo {author} {\bibfnamefont {J.~M.}\ \bibnamefont
  {Kosterlitz}},\ }\href {\doibase 10.1103/PhysRevLett.65.137} {\bibfield
  {journal} {\bibinfo  {journal} {Phys. Rev. Lett.}\ }\textbf {\bibinfo
  {volume} {65}},\ \bibinfo {pages} {137} (\bibinfo {year} {1990})}\BibitemShut
  {NoStop}%
\bibitem [{\citenamefont {Landau}\ and\ \citenamefont
  {Binder}(2013)}]{BookLandau13}%
  \BibitemOpen
  \bibfield  {author} {\bibinfo {author} {\bibfnamefont {D.}~\bibnamefont
  {Landau}}\ and\ \bibinfo {author} {\bibfnamefont {K.}~\bibnamefont
  {Binder}},\ }\href {https://books.google.de/books?id=hrIhAwAAQBAJ} {\emph
  {\bibinfo {title} {A Guide to Monte Carlo Simulations in Statistical
  Physics}}}\ (\bibinfo  {publisher} {Cambridge University Press},\ \bibinfo
  {year} {2013})\BibitemShut {NoStop}%
\bibitem [{\citenamefont {Radzihovsky}\ and\ \citenamefont
  {Lubensky}(2001)}]{Radzihovsky01}%
  \BibitemOpen
  \bibfield  {author} {\bibinfo {author} {\bibfnamefont {L.}~\bibnamefont
  {Radzihovsky}}\ and\ \bibinfo {author} {\bibfnamefont {T.~C.}\ \bibnamefont
  {Lubensky}},\ }\href {http://stacks.iop.org/0295-5075/54/i=2/a=206}
  {\bibfield  {journal} {\bibinfo  {journal} {EPL (Europhysics Letters)}\
  }\textbf {\bibinfo {volume} {54}},\ \bibinfo {pages} {206} (\bibinfo {year}
  {2001})}\BibitemShut {NoStop}%
\bibitem [{\citenamefont {Lebwohl}\ and\ \citenamefont
  {Lasher}(1972)}]{Lebwohl72}%
  \BibitemOpen
  \bibfield  {author} {\bibinfo {author} {\bibfnamefont {P.~A.}\ \bibnamefont
  {Lebwohl}}\ and\ \bibinfo {author} {\bibfnamefont {G.}~\bibnamefont
  {Lasher}},\ }\href@noop {} {\bibfield  {journal} {\bibinfo  {journal}
  {Physical Review A}\ }\textbf {\bibinfo {volume} {6}},\ \bibinfo {pages}
  {426} (\bibinfo {year} {1972})}\BibitemShut {NoStop}%
\bibitem [{\citenamefont {Gaenko}\ \emph {et~al.}(2017)\citenamefont {Gaenko},
  \citenamefont {Antipov}, \citenamefont {Carcassi}, \citenamefont {Chen},
  \citenamefont {Chen}, \citenamefont {Dong}, \citenamefont {Gamper},
  \citenamefont {Gukelberger}, \citenamefont {Igarashi}, \citenamefont
  {Iskakov}, \citenamefont {Könz}, \citenamefont {LeBlanc}, \citenamefont
  {Levy}, \citenamefont {Ma}, \citenamefont {Paki}, \citenamefont {Shinaoka},
  \citenamefont {Todo}, \citenamefont {Troyer},\ and\ \citenamefont
  {Gull}}]{Gaenko17}%
  \BibitemOpen
  \bibfield  {author} {\bibinfo {author} {\bibfnamefont {A.}~\bibnamefont
  {Gaenko}}, \bibinfo {author} {\bibfnamefont {A.}~\bibnamefont {Antipov}},
  \bibinfo {author} {\bibfnamefont {G.}~\bibnamefont {Carcassi}}, \bibinfo
  {author} {\bibfnamefont {T.}~\bibnamefont {Chen}}, \bibinfo {author}
  {\bibfnamefont {X.}~\bibnamefont {Chen}}, \bibinfo {author} {\bibfnamefont
  {Q.}~\bibnamefont {Dong}}, \bibinfo {author} {\bibfnamefont {L.}~\bibnamefont
  {Gamper}}, \bibinfo {author} {\bibfnamefont {J.}~\bibnamefont {Gukelberger}},
  \bibinfo {author} {\bibfnamefont {R.}~\bibnamefont {Igarashi}}, \bibinfo
  {author} {\bibfnamefont {S.}~\bibnamefont {Iskakov}}, \bibinfo {author}
  {\bibfnamefont {M.}~\bibnamefont {Könz}}, \bibinfo {author} {\bibfnamefont
  {J.}~\bibnamefont {LeBlanc}}, \bibinfo {author} {\bibfnamefont
  {R.}~\bibnamefont {Levy}}, \bibinfo {author} {\bibfnamefont {P.}~\bibnamefont
  {Ma}}, \bibinfo {author} {\bibfnamefont {J.}~\bibnamefont {Paki}}, \bibinfo
  {author} {\bibfnamefont {H.}~\bibnamefont {Shinaoka}}, \bibinfo {author}
  {\bibfnamefont {S.}~\bibnamefont {Todo}}, \bibinfo {author} {\bibfnamefont
  {M.}~\bibnamefont {Troyer}}, \ and\ \bibinfo {author} {\bibfnamefont
  {E.}~\bibnamefont {Gull}},\ }\href {\doibase
  https://doi.org/10.1016/j.cpc.2016.12.009} {\bibfield  {journal} {\bibinfo
  {journal} {Computer Physics Communications}\ }\textbf {\bibinfo {volume}
  {213}},\ \bibinfo {pages} {235 } (\bibinfo {year} {2017})}\BibitemShut
  {NoStop}%
\end{thebibliography}%

\end{document}